\documentclass[11pt]{article}

\usepackage{epsfig}

\def\be{\begin{equation}}
\def\ee{\end{equation}}
\def\bea{\begin{eqnarray}}
\def\eea{\end{eqnarray}}

\title{Gravitational Perturbations of Higher Dimensional Rotating Black Holes: Tensor Perturbations}

\author{Hari K. Kunduri \\ DAMTP, University of Cambridge, Wilberforce Road, Cambridge, CB3 0WA, UK \\ h.k.kunduri@damtp.cam.ac.uk \\ \\ James Lucietti \\ St. John's College, University of Oxford, Oxford, OX1 3JP, UK \\ j.lucietti@damtp.cam.ac.uk \\  \\Harvey S. Reall \\ School of Physics and Astronomy, University of Nottingham, NG7 2RD, UK \\harvey.reall@nottingham.ac.uk \\ \\ DAMTP-2006-46}
\date{8 June, 2006}

\begin{document}

\maketitle

\begin{abstract}
\noindent
Assessing the stability of higher-dimensional rotating black holes requires a study of linearized gravitational perturbations around such backgrounds. We study perturbations of Myers-Perry black holes with equal angular momenta in an odd number of dimensions (greater than five), allowing for a cosmological constant. We find a class of perturbations for which the equations of motion reduce to a single radial equation. In the asymptotically flat case we find no evidence of any instability. In the asymptotically anti-de Sitter case, we demonstrate the existence of a superradiant instability that sets in precisely when the angular velocity of the black hole exceeds the speed of light from the point of view of the conformal boundary. We suggest that the endpoint of the instability may be a stationary, nonaxisymmetric black hole.
\end{abstract}

\section{Introduction}

Exact solutions describing higher-dimensional rotating black holes have been known for a long time \cite{MP} but the question of their classical stability is still unresolved. There are arguments suggesting that a Myers-Perry (MP) black hole will be unstable for sufficiently large angular momentum in five \cite{ER} and higher \cite{EM} dimensions. However, a convincing demonstration of this requires a study of linearized gravitational fluctuations around such backgrounds. This has only been done in the limit of vanishing angular momentum, i.e., for higher-dimensional Schwarzschild black holes \cite{HG,kodama:03a,kodama:03}.

Another context in which stability of higher dimensional rotating black holes has been discussed is the AdS/CFT correspondence \cite{adscft}. The MP solutions have been generalized to include a cosmological constant \cite{HHT,GLPP}. There is a qualitative argument that rotating, asymptotically AdS black holes might exhibit a superradiant instability \cite{HR}. The idea (inspired by the corresponding instability of a Kerr black hole in the presence of a massive scalar field \cite{kerrinstab,eardley,detweiler}) is that superradiant perturbations are trapped by the AdS potential barrier at infinity and reflected towards the black hole where they get amplified and the process repeats. It can be proved that no such instability is present for black holes rotating at, or slower than, the speed of light relative to the conformal boundary \cite{HR} (essentially because superradiant modes do not fit into the AdS "box" \cite{cardosoa}) but an instability may well be present for more rapidly rotating holes. The only way to find out is to study perturbations of such black holes. This has been done for scalar field perturbations of small four-dimensional Kerr-AdS black holes \cite{cardoso} but not for gravitational perturbations, large black holes, or higher dimensions.

The main goal of this paper is to study gravitational perturbations of higher-dimensional rotating black holes.
In four dimensions, gravitational perturbations of rotating black
holes can be studied analytically. The tractability of the problem
arises from two miraculous properties of the Kerr metric. Firstly, the
equations of motion for gravitational perturbations can be decoupled
and reduced to a PDE for a single scalar quantity
\cite{teukolsky}. Secondly, this equation can be reduced to ODEs
governing the radial and angular behaviour by separation of
variables. It is known that this separability property of the Kerr
metric does extend to some of the higher-dimensional MP metrics
\cite{frolov,ida,page1}, and MP metrics with a cosmological constant~\cite{page2}. This makes the study of scalar field perturbations in such backgrounds tractable. However, so far no-one has succeeded in decoupling the equations of motion governing gravitational perturbations of MP black holes.

In this paper, we will make progress with this problem by considering
the subclass of MP black holes for which the number of space-time
dimensions is odd and the angular momenta are all equal, allowing for
a cosmological constant. Such black holes are cohomogeneity-1: the
metric depends only on a radial coordinate. In $D=2N+3$ dimensions,
the isometry group is enhanced from $R \times U(1)^{N+1}$ to $R \times
U(N+1)$ where $R$ denotes time translations. The horizon is a
homogeneously squashed $S^{2N+1}$ viewed as a $S^1$ bundle over $CP^N$. The symmetry enhancement allows us to classify gravitational pertubations into scalar, vector and tensor types according to how they transform under isometries of $CP^N$. Tensor perturbations do not exist for $N=1$ so we restrict ourselves to $N>1$, i.e., spacetime dimensionality seven or greater. In general, the different types of perturbation are coupled together but we shall show that there is a subset of tensor perturbations which decouples. After separation of variables, the equations of motion for such perturbations reduce to a single ODE governing the radial behaviour. For completeness, we also consider massive scalar field perturbations of these black holes (for $N \ge 1$). We shall present a unified form for the radial equation that applies both to scalar fields and to the gravitational perturbations just discussed.

Asymptotically flat, cohomogeneity-1 Myers-Perry black holes exhibit an upper bound on their angular momentum for a given mass. Solutions saturating this bound have a regular but degenerate horizon. This means that the black holes considered here behave rather differently from ones for which some of the angular momenta vanish, which are the ones expected to exhibit a gravitational instability \cite{ER,EM}. So there is no {\it a priori} reason to expect any instability to exist for the MP black holes considered in this paper and indeed we find no instability.

In the asymptotically anti-de Sitter case, there is also an upper bound on angular momentum for given mass and black holes saturating this bound have a regular but degenerate horizon. These extremal solutions always rotate faster than light and can be arbitrarily large compared with the AdS radius. The argument of \cite{HR} suggests that black holes near to extremality might be unstable to losing energy and angular momentum into superradiant gravitational and scalar field perturbations. We shall demonstrate that this is indeed the case. Moreover, we  shall show that this instability appears as soon as the angular velocity of the hole exceeds the speed of light, i.e., as soon as the stability argument of \cite{HR} fails. The instability is a short distance instability in the sense that unstable modes exist for all wavelengths below a certain critical value determined mainly by the angular velocity of the hole. However, amongst unstable modes, the shortest wavelength modes are the least unstable.

Having demonstrated the existence of an instability, it is natural to ask what the endpoint of the instability is. We propose that a black hole that suffers from this superradiant instability will evolve to a stationary, nonaxisymmetric black hole solution. The motivation behind our proposal will be explained at the end of this paper.

This paper is organized as follows. We start by reviewing the black
hole solutions of interest. We derive equations governing
perturbations of these solutions in section \ref{sec:perturb}. Our
strategy for analyzing stability is outlined in section
\ref{sec:stability}. The AdS case turns out to be more straightforward
than the asymptotically flat case, so we analyze stability of AdS
black holes in section \ref{sec:ads} before analyzing asymptotically
flat black holes in section \ref{sec:AF}. We discuss the implications
of our results in section \ref{sec:discuss}. We also provide an
Appendix, which contains many of the calculational details of the paper.
\\ \\
{\bf Note added:} it can be shown that a general gravitational
perturbation can be decomposed into scalars, vectors and tensors on
$CP^N$, where the tensors satisfy the "doubly transverse" condition
referred to in the text and the vectors satisfy an analogous
condition. Therefore it seems appropriate to redefine what one means
by "tensor" and "vector" to include this condition. The equations of
motion for scalar, vector and tensor perturbations then decouple
from each other without any further restrictions. The present paper
deals with the most general tensor perturbations. Further details
will appear elsewhere \cite{KLR}.

\section{The background solution}

The higher-dimensional generalization of the Kerr solution was obtained by Myers and Perry \cite{MP} and subsequently generalized to include a cosmological constant in five \cite{HHT} and higher \cite{GLPP} dimensions. It is parameterized by a mass parameter $M$ and $[(D-1)/2]$ angular momentum parameters $a_i$. In $D=2N+3$ dimensions with equal rotation parameters $a_i=a$ the solution is cohomogeneity 1.
The metric can be written as:\footnote{
This can be obtained from the metric of \cite{GLPP} in "unified" Boyer-Lindquist coordinates by
changing the radial variable $r^2 \to \frac{r^2+a^2}{\Xi}$ and $M\to
M/\Xi^{N+2}$.}
\be
ds^2 = -f(r)^2dt^2 +g(r)^2dr^2 + h(r)^2[d\psi +A_a dx^a - \Omega(r)dt]^2 + r^2 \hat{g}_{ab} dx^a dx^b
\ee
\be
  g(r)^2 = \left(1+\frac{r^2}{\ell^2} - \frac{2M\Xi}{ r^{2N}} + \frac{2Ma^2}{r^{2N+2}}\right)^{-1}, \qquad h(r)^2 = r^2\left( 1+ \frac{2Ma^2}{ r^{2N+2}} \right), \qquad \Omega(r) = \frac{2Ma}{ r^{2N} h^2},
\ee
\be
 f(r) = \frac{r}{g(r)h(r)}, \qquad \Xi = 1- \frac{a^2}{\ell^2},
\ee
where $\hat{g}_{a b}$ is the Fubini-Study metric on $CP^{N}$
with Ricci tensor $\hat{R}_{ab} =2(N+1) \hat{g}_{ab}$, and $A = A_a dx^a$ is a 1-form such that $J=\frac{1}{2} dA$ is the K\"ahler form on $CP^{N}$. This way of writing the metric arises from the fact that $S^{2N+1}$ can be written as an $S^1$ fibre over $CP^{N}$. The fibre is parameterized by the coordinate $\psi$, which has period $2\pi$.

The spacetime metric satisfies $R_{\mu\nu} =
-\ell^{-2}(D-1)g_{\mu\nu}$. Asymptotically, the solution approaches
anti-de Sitter space with radius of curvature $\ell$. An
asymptotically flat Myers-Perry black hole can be recovered by taking
$\ell \rightarrow \infty$. The event horizon located at $r=r_+$ (the
largest real root of $g^{-2}$) is a Killing horizon of $\xi =
\partial_t+\Omega_H \partial_\psi$, where the angular velocity of the
horizon is: \be \Omega_H = \frac{2Ma}{r_+^{2N+2}+2Ma^2}. \ee
The mass $E$ and angular momentum $J$
(defined with respect to $\partial_{\psi}$) are~\cite{GPP}
\be E =
\frac{A_{2N+1}}{4\pi G}M\left (N+\frac{1}{2} + \frac{a^2}{2\ell^2} \right), \qquad J  =
\frac{A_{2N+1}}{4\pi G} (N+1)Ma
\ee
where $A_{2N+1}$ is the area of a unit $2N+1$ sphere.

As written the metric is parameterised by $(M,a)$. We shall assume $a
\ge 0$, which can always be achieved by $t \rightarrow -t$ if
necessary. Sometimes it will be convenient to work with more
"physical" variables $(\Omega_H,r_+)$. Fortunately one can easily
invert for $(M,a)$ in terms of $(\Omega_H,r_+)$: \be M=
\frac{r_+^{2N}(1+r_+^2\ell^{-2})^2}{2(1+r_+^2\ell^{-2} -
r_+^2\Omega_H^2)}, \qquad a =
\frac{r_+^2\Omega_H}{1+r_+^2\ell^{-2}}. \ee  For given $r_+$, existence of a regular event
horizon imposes an upper bound on $\Omega_H$: \be
 \Omega_H \le \frac{1}{\ell} \sqrt{ 1 + \frac{N \ell^2}{(N+1) r_+^2} }.
\ee
The extremal solution saturating this bound has a regular but degenerate horizon.

In the asymptotically AdS case, the "co-rotating" Killing vector field $\xi$ is timelike everywhere outside the horizon if $\Omega_H \le 1/\ell$ but becomes spacelike in a neighbourhood of infinity otherwise. With respect to the metric on the conformal boundary, $\xi$ is timelike if $\Omega_H < 1/\ell$, null if $\Omega_H = 1/\ell$ and spacelike otherwise. For this reason, black holes with $\Omega_H>1/\ell$ are said to be rotating faster than light. Note that the extremal black holes always rotate faster than light, and that such black holes can be arbitrarily large compared with the AdS length.

\section{Perturbation equations}

\label{sec:perturb}

\subsection{Scalar perturbations}

Consider a scalar field $\Phi$ obeying the Klein-Gordon equation
\be
 \nabla^2 \Phi - \mu^2 \Phi = 0.
\ee
We take a separable Ansatz:
\be
 \Phi = e^{-i \omega t + i m \psi} R(r) Y(x),
\ee
where $m$ is an integer. It is convenient to imagine performing a Kaluza-Klein reduction along $\partial/\partial \psi$. This reduces $S^{2N+1}$ to $CP^{N}$. In the reduced description, there is a gauge field $A$ on $CP^{N}$ with field strength $2J$, and $Y$ transforms as a scalar field of charge $m$ with respect to this gauge field. We assume that $Y$ is an eigenfunction of the charged scalar Laplacian on $CP^{N}$:
\be
 -{\cal D}^2 Y = \lambda Y,
\ee
where
\be
 {\cal D}_a \equiv D_a - i m A_a
\ee
is the gauge-covariant derivative on $CP^{N}$. The Klein-Gordon equation then reduces to an equation for $R(r)$. This can be written in the form of a time-independent Schr\"odinger equation
\be
\label{eqn:schr}
-\frac{d^2}{dx^2} \Psi(x) + V(x) \Psi(x) =0,
\ee
where
\be
\label{eqn:Psidef}
 \Psi = \sqrt{h}r^{N} R,
\ee
$x$ is a "tortoise" coordinate defined by
\be
\label{eqn:xdef}
 \frac{dx}{dr} = \frac{g}{f},
\ee
and the potential $V$ is
\be
 V(x) = V_0  - (\omega-m \Omega)^2 + f^2 \left( \mu^2 + \frac{m^2}{h^2}  + \frac{\lambda}{r^2} \right),
\ee
where
\be
\label{eqn:V0def}
 V_0 = \frac{f^2\sqrt{h}}{r^{N+1}}\frac{d}{dr}\left[\frac{f^2
h}{r}\frac{d}{dr}(\sqrt{h}r^{N}) \right].
\ee

\subsection{Gravitational perturbations}

We consider a linearized metric perturbation $g_{\mu\nu} \to g_{\mu\nu} +h_{\mu\nu}$ that is transverse and traceless with respect to the background metric:
\be
 \nabla^\mu h_{\mu \nu} =  g^{\mu \nu} h_{\mu\nu} = 0.
\ee
The linearized Einstein equation is
\be
 \Delta_L h_{\mu\nu} = -2\ell^{-2}(D-1) h_{\mu\nu},
\ee
where the Lichnerowicz operator is defined by
\be
\Delta_L h_{\mu\nu} = -2R_{\mu \rho \nu \sigma} h^{\rho
  \sigma} -\nabla^2 h_{\mu \nu} -2\ell^{-2}(D-1)h_{\mu\nu}.
\ee
Perturbations can be decomposed into scalar, vector and tensor parts according to how they transform under isometries of the $CP^{N}$ base space. In general, these different types of perturbation will be coupled together because a perturbation of one type can be converted into one of a different type through the operations of differentiation and contraction with the K\"ahler form. For example, a vector quantity $V_a$ can be converted into a scalar $J^{ab} (dV)_{ab}$. However, we shall identify a particular subset of tensor perturbations that can be decoupled from everything else.

A general separable tensor perturbation takes the form
\be
h _{t \mu}=h_{r\mu}=h_{\psi \mu}=0, \qquad h_{a b} = {\rm Re} \left[e^{-i\omega t+ im
\psi} r^2 R(r) Y_{ab}(x) \right],
\ee
where $Y_{ab}(x)$ is a charge $m$ symmetric tensor on $CP^{N}$. Tracelessness of $h_{\mu\nu}$ implies that $Y_{ab}$ must be traceless on $CP^{N}$. Transversality of $h_{\mu\nu}$ is equivalent to
\be
\label{transverse}
 {\cal D}^b Y_{ab} = 0,
\ee
where ${\cal D}$ is the gauge-covariant derivative on $CP^{N}$ defined above. It is well-known that there are no tensor modes on $CP^1 = S^2$ so we shall assume $N \ge 2$ henceforth.

In Appendix \ref{app:lich}, it is shown that the linearized Einstein equation reduces to two equations:
\be
\label{Jtrans}
 J^{ab}\mathcal{D}_{a}Y_{bc}=0,
\ee
and
\bea
\label{einstein2}
 0 &=& \left[ - g^{-1} \left( g^{-1} R' \right)' - \left( \frac{f'}{f} + \frac{h'}{h} + \frac{2N}{r} \right) g^{-2} R' - f^{-2} \left( \omega - m \Omega \right)^2 R + m^2 h^{-2} R  - \frac{4(N+1)}{r^2}R  \right] Y_{ab} \nonumber \\
 &+& \left[ \frac{1}{r^2} \Delta_L^A Y_{ab} + \frac{2 i m}{r^2} [J,Y]_{ab} - \frac{4h^2}{r^4} J_{ac} \{J,Y \}_{cb} \right] R,
\eea
where a prime denotes a derivative with respect to $r$, and $\Delta_L^A$ is a gauge-covariant version of the Lichnerowicz operator on $CP^{N}$:
 \be
 \Delta_L^A Y_{ab} = -2\hat{R}_{acbd}Y^{cd} - \mathcal{D}^2 Y_{ab}
+4(N+1)Y_{ab},
\ee
where $\hat{R}_{abcd}$ is the Riemann tensor of $CP^{N}$.

Equation (\ref{Jtrans}) arises from the coupling between vectors and tensors: tensor perturbations can source vector perturbations and since we have set the latter to zero, the source must vanish, which gives equation (\ref{Jtrans}). In other words, this equation is the condition that a tensor perturbation must satisfy if it is to decouple from vector perturbations. If $Y_{ab}$ satisfies both (\ref{transverse}) and (\ref{Jtrans}) then we shall refer to it as "doubly transverse". We will demonstrate later that such tensors do indeed exist.

Define a map $\mathcal{J}$ on symmetric two-tensors by
\be
(\mathcal{J}Y)_{ab}=J_a^{\phantom{a}c}J_{b}^{\phantom{b}d}Y_{cd}.
\ee
This maps traceless tensors to traceless tensors, and has
eigenvalues $\pm 1$ whose eigenvectors we refer to as hermitian  or
anti-hermitian respectively. It also maps doubly transverse tensors to doubly transverse tensors and commutes with $\Delta_L^A$. Hence we can simultaneously diagonalize $\mathcal{J}$ and  $\Delta_L^A$ on the space of traceless doubly transverse tensors. This implies that we can classify eigenfunctions of $\Delta_L^A$ into hermitian and anti-hermitian. Further, in the antihermitian case, one
can distinguish between $Y_{ab}$ of type $(2,0)$ and $(0,2)$, which we define
by $J_a^{\phantom{a}c}Y_{cb}= \epsilon iY_{ab}$, where $\epsilon =+1,-1$ for $(2,0)$ and $(0,2)$ harmonics respectively.

We can now plug all of this into the remaining equation (\ref{einstein2}). This reduces to an equation for $R(r)$ which can be rewritten in the Schr\"odinger form (\ref{eqn:schr}) by defining $\Psi$ as in (\ref{eqn:Psidef}),
$x(r)$ by equation (\ref{eqn:xdef}),
and the potential is
\be
 V(x) = V_0 - (\omega-m \Omega)^2 + f^2 \left( \frac{m^2}{h^2} + \frac{4(1-\sigma)h^2}{r^4}
 + \frac{\lambda-4(N+1)-2\epsilon(1+\sigma)m}{r^2} \right),
\ee
where $\sigma=1$ or $-1$ for anti-hermitian and hermitian $Y_{ab}$ respectively, $\lambda$ is the eigenvalue of $\Delta_L^A$, and $V_0$ is defined by (\ref{eqn:V0def}).

\subsection{Harmonics on $CP^{N}$}

We need to know the eigenvalues of charged scalar and tensor harmonics on $CP^{N}$. Start with the scalars. For $N=1$ we have $CP^1 = S^2$. Charged scalar harmonics on $S^2$ have been studied in the context of magnetic monopoles \cite{wu:76}. The $N=2$ case was studied in \cite{warner:82}. The result for general $N$ is \cite{pope:00}
\be
 \lambda = l (l+2N) - m^2, \qquad l = 2k + |m|, \qquad k=0,1,2,\ldots
\ee
These charged scalar eigenfunctions on $CP^{N}$ lift to eigenfunctions of the scalar Laplacian on $S^{2N+1}$ where $l$ is the usual "total angular momentum" quantum number.\footnote{Note that $m=\sum_i m_i$ where $(m_1,m_2,\ldots)$ is the weight vector of ${\rm SO}(2N+2)$ corresponding to the harmonic in question.}

Recall that there are no tensor harmonics on $CP^1$. Uncharged tensor harmonics on $CP^2$ have been considered in \cite{pope:82,warner:82}. It is straightforward to generalize the analysis of \cite{pope:82} to include charge: see Appendix \ref{app:CP2} for details. The result is that doubly transverse charged tensor harmonics must be anti-hermitian ($\sigma=1$). The eigenvalue spectrum is
\be
 N=2: \qquad \lambda = l(l+4) +12 -m^2 +4 \epsilon m, \qquad l = 2k + |6 - \epsilon m|
\ee
where
\be
 k = \left\{ \begin{tabular}{lll} $0,1,2, \ldots$ & if   & $\epsilon m \le 4$ \\ $1,2,3 \ldots$ & if & $\epsilon m = 5$ \\ $2,3,4, \ldots$ & if & $\epsilon m \ge 6$ \end{tabular} \right.
\ee
This implies that the minimum value of $l$ is $l_{\rm min}=6-\epsilon m$ if $\epsilon m \le 3$ and $l_{\rm min} = \epsilon m -2$ for $\epsilon m \ge 4$. Conversely, for given $l$, the allowed values of $m$ are given by
\be
\label{eqn:mrange}
 \epsilon m = 6-l, \, 6-l+2, \, 6-l+4, \, \ldots, \, l, \, l+2, \qquad l=2,3,4, \ldots
\ee
Once again, these harmonics lift to uncharged tensor harmonics on $S^5$ with $l$ the total angular momentum quantum number. Note $l \ge 2$, as expected for tensor harmonics.

Uncharged tensor harmonics on $CP^{N}$ have been studied in \cite{boucetta} with the result that doubly transverse tensors have eigenvalues
\be
m=0: \qquad  \lambda = l(l+2N)+4N+4 \sigma,
\ee
where $l=2k$, $k=0,1,2,\ldots$ is once again the total angular momentum on $S^{2N+1}$. However, the results of \cite{boucetta} should be treated with some caution because this reference overlooks the fact that $\sigma=-1$ is not permissible for $CP^2$, and also gives an incorrect lower bound for $l$, i.e., $l \ge 0$ (whereas we know from the above that $l \ge 6$ for $m=0$ on $CP^2$). The problem with the analysis of \cite{boucetta} is that it does not investigate the circumstances under which the would-be eigenfunctions vanish.

We can obtain a general formula for the charged tensor eigenvalues by comparing our results with those for non-rotating black holes. If we set $a=0$ then we have a Schwarzschild(-AdS) black hole and our tensor perturbations should form a subset of the tensor perturbations considered in \cite{HG,kodama:03a,kodama:03}. Demanding agreement between our results and those of \cite{kodama:03} yields
\be
 \lambda = l (l+2N) + 4N  + 4\sigma  -m^2 + 2 \epsilon (1 + \sigma) m.
\ee
This is consistent with all of the above results. The limitation of this approach is that it does not tell us which values of $l$ are permitted beyond the obvious restriction $l \ge 2$.

\subsection{Unified potential}

To summarize, the equations of motion for both scalar field and (doubly transverse tensor) gravitational perturbations can be separated and reduced to a single radial equation, which can be written as
\be
 -\frac{d^2 \Psi}{dx^2} + V \Psi = 0,
\ee
or, in terms of $r$, as
\be
 -\frac{f}{g} \frac{d}{dr} \left( \frac{f}{g} \frac{d\Psi}{dr} \right) + V\Psi=0.
\ee
Using our results for the eigenvalues on $CP^{N}$ we can present
a general expression for the potential that applies to both scalar field and gravitational perturbations:
\be
\label{unifiedV}
 V = V_0 + f^2 \mu^2- (\omega - m \Omega)^2 + \frac{f^2}{r^2} \left[ l(l+2N) - m^2 \left( 1 - \frac{r^2}{h^2} \right) + 4(1-\sigma) \left(\frac{h^2}{r^2} - 1 \right) \right],
\ee
where $V_0$ is defined by (\ref{eqn:V0def}). For scalar field perturbations, $N \ge 1$, $\sigma=1$ and $l=2k+|m|$, $k=0,1,2,\ldots$. For gravitational pertubations, $N \ge 2$, $\mu=0$ and the permissible values of $\sigma,l$ are discussed above.

Note that anti-hermitian ($\sigma=1$) gravitational perturbations obey exactly the same equation as a massless scalar field. Therefore, one might think that results concerning the stability of (asymptotically flat) MP black holes against massless scalar field perturbations \cite{ida,morisawa,cardoso2} would imply stability with respect to $\sigma=1$ gravitational perturbations. However, these results concern black holes in five \cite{ida} and six \cite{morisawa,cardoso2} dimensions with a single non-vanishing angular momentum whereas we are interested in black holes in seven or more odd dimensions with all angular momenta equal and non-vanishing. It appears that scalar field perturbations of such black holes have not been considered previously. Furthermore, there {\it is} a difference between scalar field and gravitational perturbations: the lower bound on $l$ is different for the two cases. (For $N=2$, the lower bound for scalars can be either less than, or greater than, the lower bound for doubly transverse tensors, according to the value of $m$.)

\section{Stability analysis}

\label{sec:stability}

\subsection{Boundary conditions}

The horizon is located at $r=r_+$, which corresponds to $x \rightarrow - \infty$:
\be\label{xnh}
 x \sim \frac{1}{\alpha} \log \left( \frac{r-r_+}{r_+} \right),
\ee
where
\be\label{alpha}
 \alpha = \frac{r_+ (g^{-2})'(r_+)}{h(r_+)}.
\ee
At the horizon, $V \rightarrow -(\omega - m \Omega_H)^2$ so the solutions behave as $\exp\left(\pm i(\omega-m\Omega_H) x \right)$. Regularity on the future horizon requires that we choose the lower sign, so we have the boundary condition
\be
\label{eqn:Phidef}
 \Psi = \exp\left(-i(\omega-m\Omega_H) x \right) \Phi,
\ee
where $\Phi(r)$ is smooth at $r=r_+$. Substituting this into the Schr\"odinger equation and expanding around $r=r_+$ gives (choosing $\Phi(r_+)=1$)
\be
\label{eqn:Phibc}
 \Phi = 1 + \frac{V'(r_+) (r-r_+)}{\alpha^2 - 2i \alpha (\omega - m \Omega_H)} + {\cal O}\left((r-r_+)^2 \right).
\ee
In the asymptotically flat case, we have
\be
 x = r + {\cal O}\left(r^{-2N + 1}\right) \qquad {\rm as} \qquad r \rightarrow \infty.
\ee
The general solution as $r \rightarrow \infty$ is a superposition of outgoing and incoming waves (proportional to $e^{i k r}$ and $e^{-i k r}$ respectively):
\be
 \Psi \sim \sqrt{kr} \left[ Z_{{\rm out}} H^{(1)}_{l+N}(kr) + Z_{{\rm in}} H^{(2)}_{l+N} (kr) \right],  \qquad N \geq 2
\ee
where $H^{(i)}$ are Hankel functions and $k = \sqrt{\omega^2 - \mu^2}$. In the $N=1$ case, the solution is as above except the order of the Hankel functions is now $[(l+1)^2+2M(\mu^2-2\omega^2)]^{1/2}$. In the asymptotically flat case, we are mainly interested in gravitational perturbations for which $\mu=0$ and $k=\omega$.

In the asymptotically AdS case, we have $x \sim -\ell^2/r \rightarrow 0$ as $r \rightarrow \infty$. The asymptotic behaviour of the potential is
\be
 V \sim V_0  + \frac{r^2\mu^2}{\ell^2} \sim \frac{r^2}{\ell^4} \left((N+1)^2 - \frac{1}{4} + \mu^2 \ell^2 \right),
\ee
with corresponding asymptotic solutions $\Psi \sim r^{-1/2 \pm \sqrt{(N+1)^2 + \mu^2 \ell^2}}$. Choosing the normalizable solution corresponds to the boundary condition
\be
\label{eqn:Psiasympads}
 \Psi \sim r^{-1/2-\sqrt{(N+1)^2 + \mu^2 \ell^2}} \qquad {\rm as} \, \, \, r \rightarrow \infty.
\ee
For stability of the AdS background we demand that a scalar field obeys the Breitenl\"ohner-Freedman bound \cite{BF}
\be
 \mu^2 \ge -\frac{(N+1)^2}{\ell^2}.
\ee
A linearized instability of the black hole would correspond to a solution of the radial equation that is regular on the future horizon and vanishing at infinity, with ${\rm Im}(\omega)>0$. In the asymptotically flat case this requires $Z_{\rm in}=0$. Note that such a solution vanishes exponentially at the horizon.

\subsection{The case $m=0$}

It is easy to show that there can be no instability (whether asymptotically flat or asymptotically AdS) when $m=0$ and $\mu^2 \ge 0$.
If $m=0$ then it is natural to consider the potential
$\tilde{V}=V+\omega^2$, which does not depend on $\omega$. The
radial equation becomes
 \be
 \label{eqn:meq0}
 -\frac{d^2}{dx^2} \Psi(x) +
\tilde{V}(x) \Psi(x) =\omega^2\Psi(x) \qquad (m=0).
\ee
Assume that $\Psi$ describes an unstable mode, so ${\rm Im}(\omega)>0$ and $\Psi$ vanishes at the horizon and at infinity as described above. The differential operator on the LHS of (\ref{eqn:meq0}) is self-adjoint on such functions and hence $\omega^2$ must be real so $\omega$ is pure imaginary and $\omega^2$ is negative. Now consider the behaviour of $V_0$. Letting $X=(r/h)(d/dr)(\sqrt{h}r^N)$, we have
\be
 V_0 = \frac{f^2 \sqrt{h}}{r^{N+1}} \frac{d}{dr} \left[ g^{-2} X \right] = \frac{f^2 \sqrt{h}}{r^{N+1}}\left( X \frac{d}{dr} g^{-2} + g^{-2} \frac{d}{dr}X \right),
\ee
It is easy to show that $X>0$ and $dX/dr>0$ for any $N \ge 1$. It is also easy to see that $(d/dr) g^{-2} >0$ outside the horizon. Hence $V_0$ is positive. If we assume $\mu^2 \ge 0$ then the remaining terms in $\tilde{V}$ are non-negative so $\tilde{V}$ is positive.  Hence $\omega^2$ must be positive, which is a contradiction.

For AdS black holes, it would be interesting to see if this conclusion could be extended to tachyonic scalars satisfying the Breitenlohner-Freedman bound by combining our argument with that of \cite{BF}.

\subsection{Strategy}

We can look for unstable modes using the strategy adopted by Press and Teukolsky for the Kerr black hole \cite{PT2}. We expect the black hole to be stable for small angular momentum because we know that the higher-dimensional Schwarzschild black hole is stable \cite{kodama:03}. Hence, for small angular velocity, the only admissible solutions of the radial equation must have negative imaginary part, i.e., they are quasi-normal modes. If an instability is to appear as the angular velocity increases then one of these quasi-normal modes must cross the real axis in the complex $\omega$ plane.\footnote{Various mathematical subtleties such as modes coming in from infinity might invalidate this statement but such subtleties do not occur for Kerr \cite{hartle} and we shall ignore this possibility here.} Hence we expect the onset of instability to be indicated by the appearance of a real frequency mode at a critical value of the angular velocity. The strategy is to look for such a mode. By continuity it must have $Z_{\rm in}=0$ in the asymptotically flat case, i.e., it must be purely outgoing at infinity. In the AdS case it must obey the "normalizable" boundary condition (\ref{eqn:Psiasympads}).

Note that the radial equation is invariant under $\omega \rightarrow -\omega$ and $m \rightarrow -m$ and, for tensors, $\epsilon \rightarrow -\epsilon$.\footnote{This just corresponds to complex conjugation of the solution. This invariance arises from a discrete symmetry of the background which, in the coordinates of \cite{GLPP} is $t \rightarrow -t$, $\phi_i \rightarrow -\phi_i$. In our coordinates this amounts to $t \rightarrow -t$, $\psi \rightarrow -\psi$ and $A \rightarrow -A$.}
Hence we can assume $\omega \ge 0$. Following Press and Teukolsky \cite{PT3}, consider the Wronskian of $\Psi$ and $\bar{\Psi}$ for real $\omega$. This must be constant so we obtain
\be
 {\rm Im} \left[ \bar{\Psi} \partial_x \Psi \right]^{x_1}_{x_2}=0,
\ee
for any $x_1$ and $x_2$. Taking $x_1$ at the horizon and $x_2$ at infinity and using the boundary condition at the horizon and infinity gives, for the (massless) asymptotically flat case,
\be
\label{eqn:wronsk}
 \left(m \Omega_H - \omega \right) = \frac{2\omega}{\pi} \left( |Z_{\rm out}|^2 - |Z_{\rm in}|^2 \right).
\ee
Hence a purely outgoing mode must have
\be
 0 \le \omega \le m \Omega_H \qquad {\rm (asymptotically \, \, flat)}.
\ee
In other words, the mode must be superradiant. In the AdS case, the LHS is unchanged but the term at infinity vanishes and we must have
\be
 \omega = m \Omega_H \qquad {\rm (asymptotically \, \, AdS)}.
\ee
Physically, this is simply the statement that there cannot be a constant flux of radiation through the horizon if the flux at infinity vanishes. Note that for both cases, we must have $m \ge 0$ since $\Omega_H \ge 0$.

We shall discuss the asymptotically AdS case first because the fact that we only have to consider a single value of $\omega$ makes this case simpler to analyse than the asymptotically flat case, for which we have to consider a range of values for $\omega$.

\section{Asymptotically anti-de Sitter black holes}

\label{sec:ads}

\subsection{Form of the potential and behaviour of solutions}

Set $\omega = m \Omega_H > 0$. The potential vanishes at the horizon and is monotonically increasing just outside the horizon. For large $r$, the potential increases\footnote{If $\mu^2$ is close to the Breitenlonher-Freedman bound then the coefficient of proportionality is negative but we shan't worry about this and our results for small black holes suggest that it doesn't change the qualitative behaviour of solutions.} proportional to $r^2$. What happens in between depends on the values of the parameters. For small $\omega$, $V$ is positive everywhere. However, for sufficiently large $\omega$, there is a "classically allowed"\footnote{
Of course, everything we are doing is classical but since we have written the radial equation in the form of a Schr\"odinger equation, we can borrow terminology such as "classically allowed", "bound state" etc. from quantum mechanics.} region in which $V$ is negative. In more detail: $V$ has roots at $r=r_1,r_2$. The potential is positive for $r_+<r<r_1$, negative for $r_1<r<r_2$ and positive for $r>r_2$, i.e., there is a potential barrier separating the classically allowed region where $V$ is negative from the horizon.

Note that, in the AdS case considered here, the initial data $\Psi(r_+)$ and $\Psi'(r_+)$ are real and positive (since $\omega=m\Omega_H$). Hence $\Psi$ is real everywhere. It is easy to see that $\Psi$ will simply increase monotonically if $V$ is positive everywhere. Hence we need $V$ to be negative somewhere for an acceptable solution of the radial equation to exist. The solution $\Psi$ will increase monotonically in the potential barrier, oscillate in the classically allowed region, and then match onto a sum of growing and decaying\footnote{More precisely: non-normalizable and normalizable.} terms at large $r$. We need to tune $\Omega_H$ until the coefficient of the growing mode vanishes, i.e., until we obtain a "bound state" solution of the radial equation.

\subsection{Small AdS black holes}

Consider the case $r_+ \ll \ell$. For $r \gg r_+$, the presence of the black hole does not affect the potential, which is determined entirely by the AdS background.
For $r \gg \ell$ the potential increases as $r^2$ and for $r_+ \ll r \ll \ell$, there is a centrifugal barrier in which $V \sim 1/r^2$. This matches onto the decreasing section of the potential following its local maximum at $r \sim r_+$. If $\omega=m\Omega_H$ is large enough then $V$ will be negative at its local minimum, which occurs at $r \sim \ell$.

The solution in the region $r \gg r_+$ is easy to obtain since we can approximate it by a solution in pure AdS. The solution obeying "normalizable" boundary conditions as $r \rightarrow \infty$ is
\be
\label{eqn:hypergeom}
 \Psi = r^{l + N+1/2}\left(1 + \frac{r^2}{\ell^2}\right)^{\omega \ell/2-a}  F\left(a,c-b;a-b+1;\frac{1}{1+r^2/\ell^2} \right),
\ee
where $F$ is the hypergeometric function, and
\bea
 \nonumber a &=& \frac{\omega \ell + l + N+1 + \sqrt{(N+1)^2 + \mu^2 \ell^2}}{2}, \qquad b = \frac{\omega \ell + l + N+1 - \sqrt{(N+1)^2 + \mu^2 \ell^2}}{2}, \\ c &=& l + N+1.
\eea
Now pick $r_0$ so that $r_+ \ll r_0 \ll \ell$. Near $r=r_0$, the solution (\ref{eqn:hypergeom}) is a superposition of a growing solution $\sim r^{l+N+1/2}$ and a decaying solution $\sim r^{1/2-l-N}$. The latter will dominate for sufficiently small $r_0$ ($r_0$ can be made as small as we like by decreasing $r_+$). Hence the solution will be monotonically decreasing near $r=r_0$. However, $r=r_0$ is inside the potential barrier outside the horizon of the black hole, where we know that $\Psi$ should be increasing! So in fact the coefficient of the decreasing solution must vanish, which leads to the quantization rule $c-b=-p$ where $p$ is a non-negative integer. This implies $\omega \ell = l+N+1+\sqrt{(N+1)^2 + \mu^2 \ell^2} + 2p$. This is just the usual quantization of frequency in global AdS.
However, here we know that $\omega = m \Omega_H$. Hence, for small AdS black holes, a real frequency solution to the radial equation exists for
\be
\label{eqn:smallresult}
 \Omega_H \ell = \frac{l+N+1+\sqrt{(N+1)^2 + \mu^2 \ell^2} + 2p}{m}, \qquad p=0,1,2,\ldots
\ee
The solutions with $p>0$ correspond to "excited states" for which the solution of the radial equation oscillates (with $p+1$ extrema) before approaching zero at large $r$. We are interested in the onset of instability, corresponding to the smallest value of $\Omega_H \ell$ for which a solution exists, so we are mainly interested in $p=0$.

For scalars, we have $l \ge m$ so, for given $m$, the smallest value of $\Omega_H$ for which we have a solution is
\be
 \Omega_H \ell = 1 + \frac{N+1 + \sqrt{(N+1)^2+\mu^2 \ell^2}}{m}.
\ee
This is the critical value of $\Omega_H$ beyond which modes with angular quantum number $m$ become unstable. Note that it always exceeds $1/\ell$, consistent with the proof of stability for $\Omega_H \le 1/\ell$ given in \cite{HR}. However, this proof has been criticized \cite{GM} because it assumes the dominant energy condition, which is violated if $\mu^2<0$. Our result shows that, for small black holes, this does not matter so long as the Breitenl\"ohner-Freedman bound is satisfied.

Note that the critical value for $\Omega_H$ tends to $1/\ell$ from above as $m \rightarrow \infty$. This proves that, for small black holes at least, the instability sets in as soon as $\Omega_H$ exceeds $1/\ell$, with the shortest wavelength modes becoming unstable first.

Now consider gravitational pertubations, for which the threshold of stability occurs at
\be
 \Omega_H \ell = \frac{l + 2N+2}{m}.
\ee
For $N=2$, taking $l=l_{\rm min}(m)$ and $\epsilon=1$, this evaluates to $11,5,3$ for $m=1,2,3$ and $1+4/m$ for $m \ge 4$. (Taking $\epsilon=-1$ just makes $l_{\rm min}$ bigger.) So the conclusion is the same as for scalar field perturbations: a superradiant gravitational instability sets in as soon as the angular velocity exceeds the speed of light, with the shortest wavelength modes becoming unstable first.

In the context of the AdS/CFT correspondence, we have to consider a higher dimensional spacetime consisting of the product of the black hole space-time with a compact internal space (e.g. a sphere). It is believed that small AdS black holes are unstable with respect to the Gregory-Laflamme instability \cite{GL} under which they are expected to localize on the internal space. This means that, although small black holes with $\Omega_H \ell \le 1$ do not suffer from a superradiant instability, they are nevertheless unstable. In order to eliminate the GL instability we have to extend our results to large AdS black holes, i.e., $r_+ > \ell$. This will be addressed in the next section.

\subsection{Numerical analysis}

Fix $r_+$ and $(l,m)$. For given $\Omega_H$ we integrate numerically the equation for $\Psi$ (with $\omega=m\Omega_H$) out from $r=r_+$. For small $\Omega_H \ell$ we find $\Psi \rightarrow \infty$ as $r \rightarrow \infty$. However, when a critical value of $\Omega_H \ell$ is exceeded, we find $\Psi \rightarrow -\infty$ as $r \rightarrow \infty$. The critical value corresponds to a solution for which $\Psi \rightarrow 0$ as $r \rightarrow \infty$, which is what we are looking for. The precise value of $\Omega_H \ell$ at which this occurs can be obtained accurately by interval bisection. This can then be repeated for different values of $r_+,l,m$.
\begin{figure}
\centering{\psfig{file=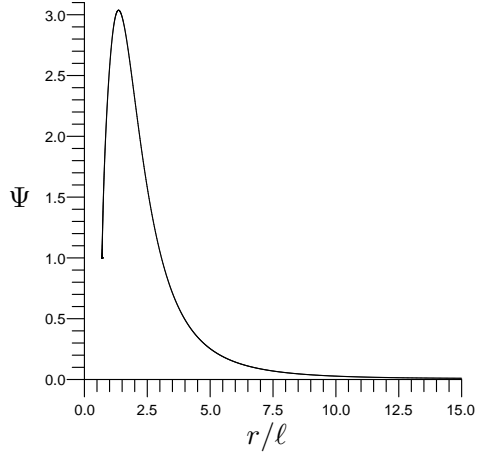,width=6cm}}
\begin{picture}(0,0)(180,0)
\put(0,90){$\Psi$}
\put(90,0){$r/\ell$}
\end{picture}
\caption{Plot of solution $\Psi(r)$ against $r/\ell$ for doubly transverse tensor perturbations with $N=2$, $l=6$, $m=8$, $r_+/\ell=0.7$, $\Omega_H\ell = 1.46242$.}
\label{fig:wavefn}
\end{figure}

\begin{figure}
\centering{\psfig{file=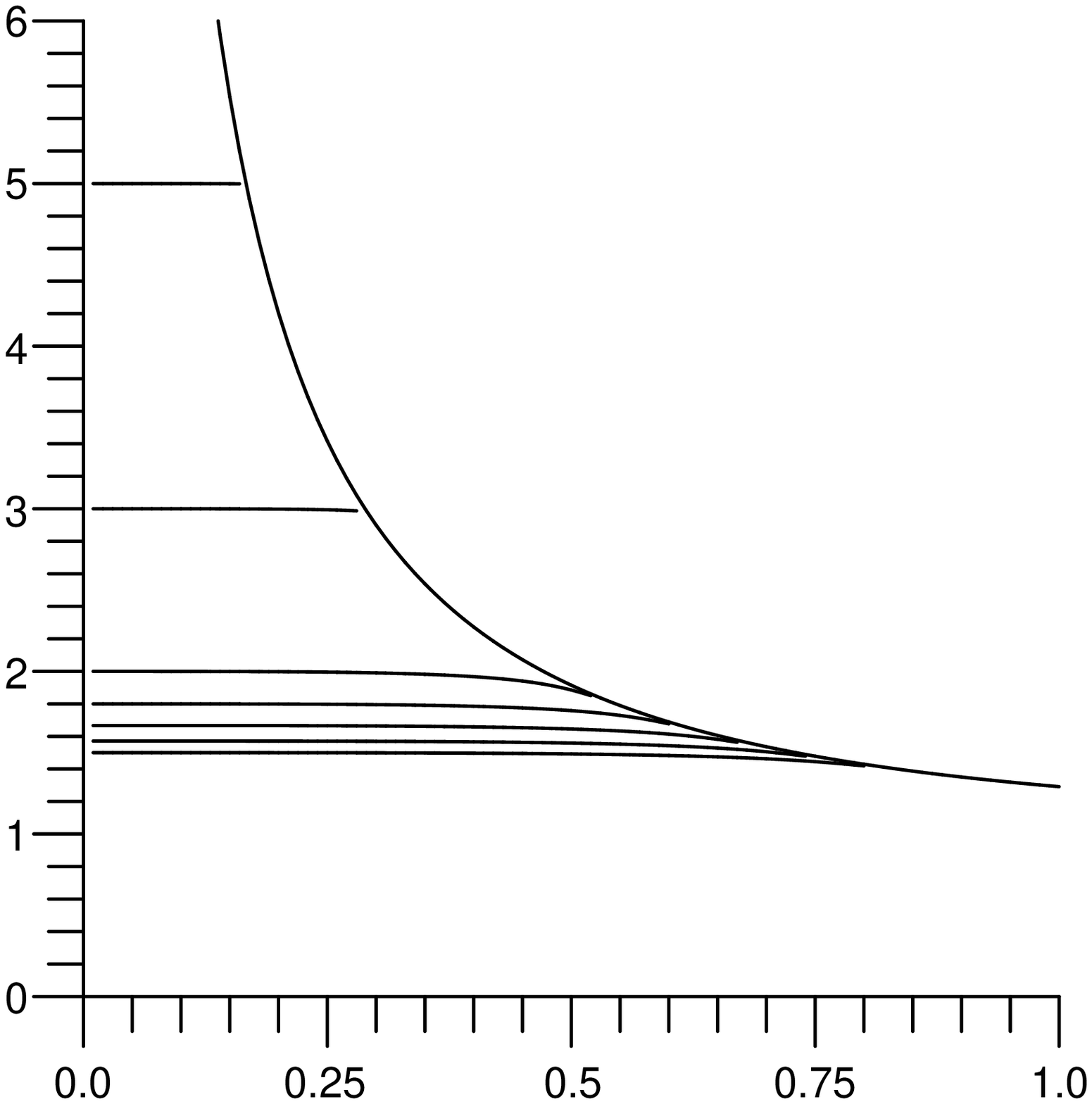,width=8cm}}
\begin{picture}(0,0)(230,0)
\put(0,120){$\Omega_H \ell$}
\put(120,10){$r_+/\ell$}
\end{picture}
\caption{Plot of $\Omega_H\ell$ against $r_+/\ell$ for doubly transverse tensor perturbations with $N=2$, $\epsilon=1$. No black holes exist in the empty region bounded by the curve in the top right of the diagram. Curves $\Omega_H(r_+,l,m)$ corresponding to existence of a normal mode with frequency $\omega = m \Omega_H$ are displayed for different values of $l,m$. From top to bottom, $(l,m) = (5,1),(4,2),(3,3),(2,4),(3,5),(4,6),(5,7),(6,8)$.}
\label{fig:eigenvalueplot}
\end{figure}

In this section, we shall consider only gravitational perturbations with $N=2$ ($D=7$). A typical numerical solution for the wavefunction at a critical value of $\Omega_H$ is shown in figure \ref{fig:wavefn}. Results for the lowest values of $m$ are shown in figure \ref{fig:eigenvalueplot}. We plot the critical value of $\Omega_H$ as a function of $r_+$ for different values of $m$ with $l=l_{\rm min}(m)$ and $\epsilon=1$. These curves terminate when they hit the curve corresponding to the extremal black hole (beyond which no black hole solution exists).
There are two points to note about these results. First, as $r_+/\ell \rightarrow 0$, they are in excellent agreement with the theoretical predictions for small black holes discussed above. Second, the only effect of increasing $r_+$ is a very slight decrease in $\Omega_H$. This decrease was too small for our numerics to detect for $m=1,2$.  For $(l,m)=(3,3)$, the critical value of $\Omega_H$ decreases from $3.00$ to $2.99$ as $r_+$ increases from $0$ to its maximum. For $(l,m)=(2,4)$ the decrease is from $2.00$ to $1.85$ and for $(l,m)=(6,8)$ it is from $1.50$ to $1.42$.

\begin{figure}
\centering{\psfig{file=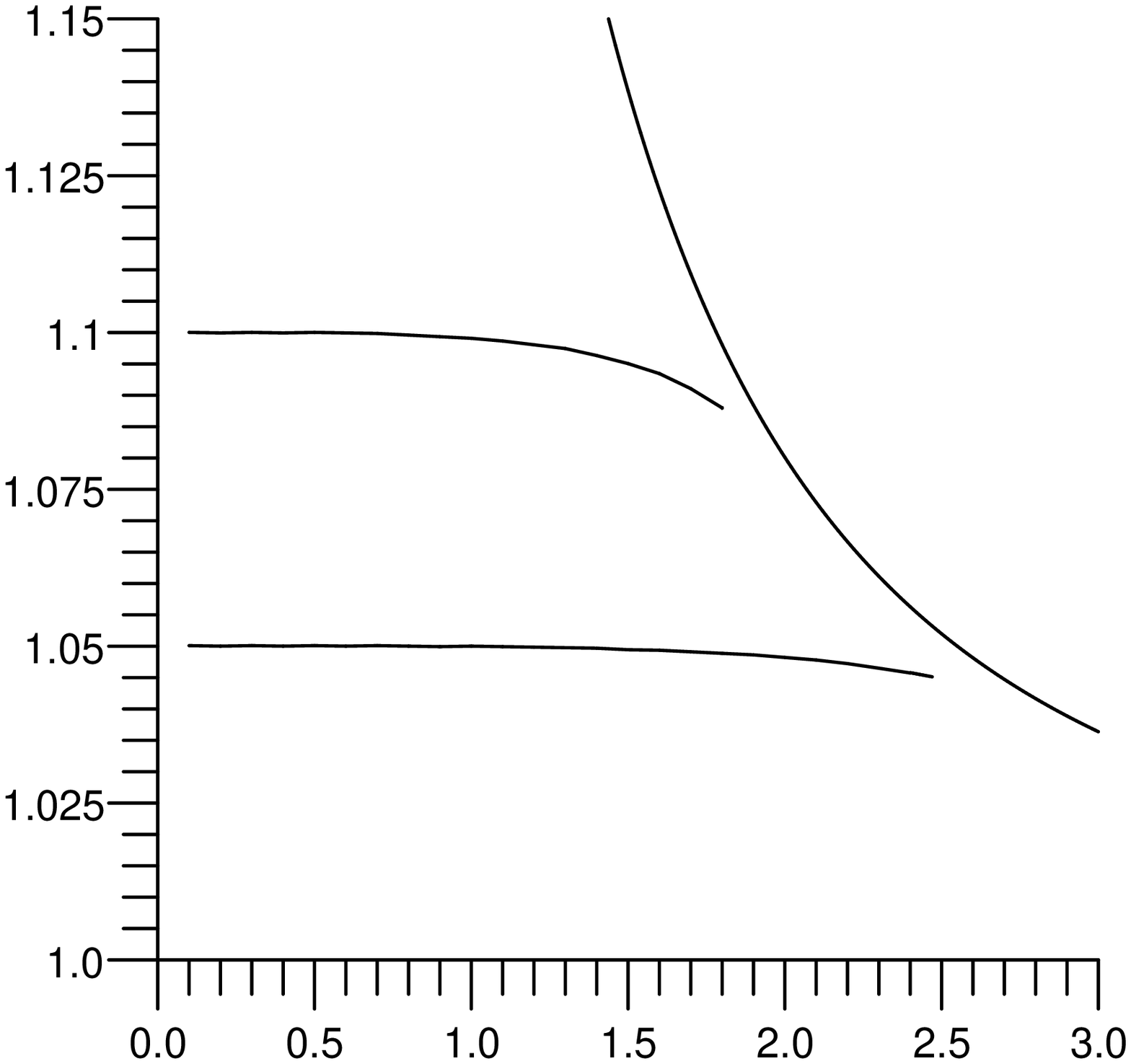,width=8cm}}
\begin{picture}(0,0)(240,0)
\put(0,120){$\Omega_H \ell$}
\put(120,10){$r_+/\ell$}
\end{picture}
\caption{Plots of critical value of $\Omega_H$ for doubly transverse tensor perturbations for $N=2$, $\epsilon=1$, $(l,m)=(38,40)$ (top) and $(78,80)$. No black holes exist in the empty region bounded by the curve in the top right of the diagram. The fact that the $\Omega_H$ curves do not quite meet this curve is due to the limitations of our numerical method. Note the scale of the vertical axis.}
\label{fig:eigenvalueplot2}
\end{figure}

To obtain results for $r_+/\ell > 1$ we have to go to higher $m$. Some sample results are presented in figure \ref{fig:eigenvalueplot2}.
Our numerical results suggests that the critical value of $\Omega_H \ell$ is always greater than $1$, and tends to $1$ as $m \rightarrow \infty$, i.e., the curves on the figures accumulate at $\Omega_H \ell \rightarrow 1+$ as $m \rightarrow \infty$. The interpretation is exactly the same as for small black holes:
for any given $r_+$, if we start from $\Omega_H=0$ and increase $\Omega_H$ then as soon as $\Omega_H\ell$ exceeds $1$, the black hole will become unstable to all perturbations for which $m$ exceeds some critical value. In the next section we shall demonstrate this analytically.

\subsection{WKB analysis}

In this section we consider both scalar field
and gravitational perturbations governed by the effective
Schr\"odinger equation with potential (\ref{unifiedV}). As before, we are
interested in modes at the threshold of instability so $\omega=m\Omega_H$. The strategy is to look
at the potential for large $m$ (and hence large $\omega$). In this limit, we
can use WKB techniques.

As discussed above, if an appropriate solution of the radial equation is to exist then the qualitative behaviour of the potential must be as follows: it
vanishes at the horizon $r=r_+$, increases to a local maximum and
then decreases and becomes negative in some range $r_1 < r < r_2$. For $r>r_2$ the potential is positive and monotonically increasing.

We are interested in bound states, for which $\Psi$ obeys the boundary condition (\ref{eqn:Psiasympads}). The appropriate WKB solution in the region $r > r_2$ is
\begin{equation}
\Psi(x) \approx \gamma^{-{1}/{2}}\exp\left( -\int_{x_2}^{x} \gamma
dx \right )
\end{equation} where $\gamma(x) = \sqrt{V(x)}$. Note that, for large $r$, we have
\be
 V(x) \approx \frac{\kappa^2}{x^2}, \qquad x \rightarrow 0-
\ee
where
\be
 \kappa =  \sqrt{(N+1)^2-\frac{1}{4} +
 \mu^2\ell^2}.
\ee
The WKB approximation is valid when $|\gamma^{-2} d\gamma/dx|$ and $|\gamma^{-3} d^2\gamma/dx^2|$  are small, which requires $\kappa \gg 1$. Hence we do not expect our results to be reliable for a scalar field with mass close to the Breitenl\"ohner-Freedman bound. For gravitational perturbations $\mu=0$ so {\it a priori} we might not expect WKB to work very well. However, the same remark applies to the calculation of black hole quasi-normal modes, where WKB has been found to be accurate \cite{schutz}. We shall see that the WKB results are in good agreement with our numerical results for $\mu=0$. In any case, the WKB method is certainly reliable for sufficiently massive scalar fields.

The standard connection formula (see, e.g.~\cite{dicke}) imply
that, in the classically allowed region $r_1 < r <r_2$, we
have \begin{equation} \Psi \approx 2k^{-{1}/{2}}\cos
\left(\int_{x}^{x_2} k dx - \frac{\pi}{4} \right )
 = 2k^{-{1}/{2}}\cos \left( I - \int_{x_1}^{x} kdx -
 \frac{\pi}{4} \right).
 \end{equation} Here $k(x) = \sqrt{-V(x)}$ and
 \begin{equation}
 I = \int_{x_1}^{x_2} k dx.
 \end{equation} Applying the connection formulae once again leads
 to the WKB solution for $r<r_1$: \be \label{eqn:wkbregion1}
 \Psi \approx 2 \gamma^{-1/2} \cos I \exp \left( \int_{-\infty}^{x_1} \gamma dx - \int_{-\infty}^x \gamma dx \right).
\ee
The WKB approximation breaks down as $x \rightarrow -\infty$ ($r \rightarrow r_+$) because $V \rightarrow 0$. This requires
separate analysis. Using~(\ref{xnh}) and (\ref{alpha}) and setting $y=\exp(\alpha x/2)$, the Schrodinger equation becomes
\be
 y^2 \frac{d^2 \Psi}{dy^2} + y \frac{d\Psi}{dy} - \left( \frac{4 r_{+} V'(r_+)}{\alpha^2} y^2 + {\cal O}(y^4) \right) \Psi = 0.
\ee This is Bessel's equation for $y \rightarrow 0$. The solutions
are $I_0(z)$ and $K_0(z)$ where \be
 z = \frac{2}{\alpha} \sqrt{r_{+}V'(r_+)} \, y.
\ee As $m$ is large,  $V'$ is large and hence for moderate $y$,
the argument of the Bessel functions is large and we can
approximate them by their asymptotic forms: \be
 I_0 (z) \sim \frac{e^z}{\sqrt{2\pi z}}, \qquad K_0(z) \sim \sqrt{\frac{\pi}{2z}} e^{-z}.
\ee In general we have a linear combination of these two
solutions. However, if $\Psi$ is asymptotically decaying for large
$z$ then it must be proportional to $K_0$. This is precisely what
we have found from our WKB analysis: equation
(\ref{eqn:wkbregion1}) decays for large $z$. Hence our WKB
solution matches onto the $K_0$ solution near the horizon.
However, the $K_0$ solution diverges at the horizon, and the only
way to avoid this is if $\cos I = 0$, i.e., \be \label{eqn:Iquant}
 I = \left( p + \frac{1}{2} \right) \pi, \qquad p = 0,1,2,\ldots
\ee
The next step is to evaluate
\be
 I = \int_{r_1}^{r_2} \sqrt{-V(r)} \frac{g}{f} dr.
\ee
We assume (motivated by our numerical results and results for small black holes) that $\Omega_H \ell = 1 + {\cal O}(1/m)$ and that the roots $r_1$ and $r_2$ are of order $\sqrt{m}
\ell$.  These assumptions can be justified {\it a posteriori}. We take $m \gg
1$, $m \gg r_+^2/\ell^2$. Then within the integral $r$ is of order
$\sqrt{m} \ell$ and we can approximate $g/f \approx \ell^2/r^2$.
Now consider~(\ref{unifiedV}) with the above restrictions.
We find
\begin{equation}
V(r) \approx \frac{\kappa^2}{\ell^4} r^2 + m^2 \left( \ell^{-2} -
\Omega_H^2 + r^{-2} \right) + \frac{2m(l-m + N)}{\ell^2}.
\end{equation}
We assume $l \sim l_{\rm min} = m + {\cal O}(1)$ for large $m$. Note that we must have $\Omega_H \ell > 1$ for $V$ to be negative somewhere. The roots $r_1$ and $r_2$ are determined by $V(r_i)=0$, with solutions
 \be
 2\kappa^2 \frac{r^2}{\ell^2} = \left[ m^2 \left( \Omega_H^2 \ell^2 -1 \right) - 2m(l-m+N) \pm \sqrt{ \left(m^2 \left( \Omega_H^2 \ell^2 -1 \right)-2m(l-m+N) \right)^2 - 4\kappa^2 m^2 } \right].
\ee We then find \be
 I = \kappa \int_{r_1}^{r_2} \frac{dr}{r^3} \sqrt{(r^2-r_1^2)(r_2^2-r^2)} = \frac{\kappa\pi (r_2-r_1)^2}{4 r_1
 r_2}, \ee where the integral was calculated by changing variables to $w=r^2$ and working in the complex $w$-plane.\footnote{More precisely, introduce a branch cut between $r_{1}^2$ and $r_{2}^2$; the integral around the branch cut can be deformed and gets contributions from a double pole at $w=0$ and the circle at infinity.} Plugging this into the quantization condition (\ref{eqn:Iquant})
gives, at large $m$ with $l-m = {\cal O}(1)$, \be
\label{eqn:wkbanswer}
 \Omega_H \ell \approx 1 + \frac{1}{m} \left(l-m+N+1 +\sqrt{(N+1)^2-\frac{1}{4} + \mu^2\ell^2} + 2p \right).
\ee Finally we can use these values for $\Omega_H$ to verify that
$r_i$ are really of order $\sqrt{m} \ell$, as previously assumed. Recall that validity of the WKB approximation requires that $(N+1)^2 + \mu^2 \ell^2 \gg 1/4$ so this result agrees with our result (\ref{eqn:smallresult}) for small black holes. Note that our result is independent of $r_+$, which explains (at least for large $m$) why our numerical results showed little variation with $r_+$. Setting $\mu=0$, $N=2$ and $p=0$, the WKB method gives $\Omega_H \ell \approx 1 + 3.96/m$ in good agreement with our numerical results so, as advertised above, the WKB method works well even though one might question its validity when $\mu=0$.

The coefficient of $1/m$ in (\ref{eqn:wkbanswer}) is positive (at least in all cases for which we know $l_{\rm min}$) so once again we see $\Omega_H \ell \rightarrow 1+$ as $m \rightarrow \infty$ so, irrespective of the size of the black hole, once its angular velocity exceeds the speed of light it becomes unstable to perturbations of arbitrarily short wavelength.

\subsection{WKB calculation of unstable modes}

So far, we have been looking for real frequency modes, whose existence indicates the onset of instability. In this section, we shall look for unstable modes directly using the WKB approach. This has been used previously in a study of the superradiant instability of the Kerr black hole in the presence of a massive scalar field \cite{eardley}.

We allow $\omega$ to be complex: $\omega=\omega_R +  i \omega_I$ and assume $|\omega_I| \ll \omega_R = {\cal O}(m)$. Neglecting $\omega_I$, the potential is now negative at $r=r_+$ and has an additional zero at $r=r_0$, $r_+<r_0 < r_1$. The potential is negative in a region $r_+ \le r<  r_0$ near the horizon, positive in the potential barrier region $r_0 < r < r_1$, negative for $r_1 < r < r_2$ and positive for $r>r_2$.
It is straightforward to extend the WKB analysis to cover this potential so we shall neglect the details. Assuming $\omega_R < m \Omega_H$ and $l-m = {\cal O}(1)$ we obtain the quantization condition
\be
 \frac{\omega_R \ell}{m} \approx1 + \frac{1}{m}\left(  l-m+N+1 +\sqrt{(N+1)^2-\frac{1}{4} + \mu^2\ell^2} + 2p \right), \qquad p=0,1,2, \ldots
\ee
The WKB solution for $r>r_1$ takes the same form as before. For $r_0<r<r_1$ the solution is ($\gamma=\sqrt{V}$ again)
\be
 \Psi \approx (-)^p \gamma^{-1/2} \exp\left(-\int_{x_0}^{x_1} \gamma dx\right) \exp \left( \int_{x_0}^x \gamma dx \right).
\ee
For $r_+ < r < r_0$ the solution is ($k = \sqrt{-V}$)
\be
 \Psi \approx (-)^p k^{-1/2} \exp\left(-\int_{x_0}^{x_1} \gamma dx\right) \exp \left( -i\pi/4 - i \int_x^{x_0} k dx \right).
\ee
To calculate $\omega_I$, write the Schr\"odinger equation as
\be
 -\partial_x^2 \Psi - \left(\omega -m \Omega \right)^2 \Psi + \hat{V} \Psi = 0,
\ee
where $\hat{V}$ is independent of $\omega$. Multiply this equation by $\bar{\Psi}$, take the imaginary part and integrate (wrt $x$) from $r=r_*>r_+$ to $r=\infty$ to obtain
\be
 2 \omega_I \int_{r=r_*}^{r=\infty} dx \left(\omega_R - m \Omega \right) |\Psi|^2 = {\rm Im} \left[\bar{\Psi}\partial_x \Psi \right]_{r=r_*}.
\ee
Take $r_*<r_0$ and evaluate the RHS using the WKB solution. The integral on the LHS is dominated by the classically allowed region $r_1 < r < r_2$ where $\omega_R \gg m \Omega$. We obtain
\be
 \omega_I \approx \frac{ \exp\left(- 2 \int_{x_0}^{x_1} \gamma dx\right) }{ 2 \omega_R \int_{x_1}^{x_2} |\Psi|^2 dx }.
\ee
This is the same as equation 41 of \cite{eardley}. The denominator can be calculated using the WKB solution and the expression for $\omega_R$. It turns out to be ${\cal O}(1)$ at large $m$. In the numerator, there is a contribution of ${\cal O}(m)$ to the integral coming from the region near $r=r_0$ so we can bound
\be
 0< \omega_I \ell < \alpha \exp (-\beta m)
\ee
for some positive constants $\alpha$, $\beta$. This justifies our assumption $\omega_I \ll \omega_R$. We see that, although large $m$ modes are the first to become unstable when $\Omega_H \ell$ exceeds $1$, the growth time of the instability is exponentially large in $m$ so these modes are the least unstable. This suggests that the most unstable modes will be those for which $m$ is not particularly large. It would be interesting to calculate $\omega_I$ for such modes.

\section{Asymptotically flat black holes}

\label{sec:AF}

\subsection{Introduction}

We know that a mode at the threshold of instability must obey $0 \le \omega \le m \Omega_H$, i.e., it is superradiant. The only known way that superradiant modes can actually lead to an instability is if they can be trapped by the potential at infinity, i.e., they must be bound states. This would require a local minimum in $V$, as in the AdS case, or for a massive scalar field in four dimensions.\footnote{
In higher dimensions, it appears that even a mass term for a scalar is not enough to lead to a superradiant instability, at least for MP black holes with a single non-vanishing angular momentum \cite{cardoso3}.} In all cases that we have examined,
the qualitative form of the potential for $\mu=0$ is: $V \rightarrow -(\omega - m \Omega_H)^2$ as $x \rightarrow -\infty$, then $V$ increases to a positive maximum and decreases to $-\omega^2$ as $x \rightarrow \infty$. It appears that a local minimum in $V$ is not possible so there is no obvious sign of any gravitational instability apparent from our radial equation.

This qualitative argument is no substitute for a quantitative study. We shall analyse the radial equation in two cases: first for large $m$ and $\omega$ using the WKB method and then numerically for $D=7$.

\subsection{WKB approximation}

Consider large $m$ with $r_+ \omega/m$ fixed and $l \sim l_{\rm min}$ so $l/m \rightarrow 1$. We find
\be
 W \equiv \lim_{m \rightarrow \infty} \frac{V}{m^2} = - \left( \frac{\omega}{m} - \Omega \right)^2 + \frac{f^2}{h^2},
\ee
This is true even with a mass term (as we're considering a limit in which $\omega \gg \mu$). The explicit form of $W$ is
\be
 W = \frac{P(r)}{r^{2N+2} + 2Ma^2},
\ee
where
\be
\label{eqn:Pdef}
 P(r) = -\frac{\omega^2}{m^2} r^{2N+2} + r^{2N} - 2M \left(1 - \frac{a\omega}{m} \right)^2.
\ee
The qualitative behaviour of $P(r)$ is increasing from a negative value at $r=r_+$ to a maximum at $r^2 = Nm^2/((N+1)\omega^2)$, then decreasing to $-\infty$ as $r \rightarrow \infty$. The important question is whether the maximum of $P$ is positive or negative. If it were negative then $V$ would be everywhere negative so the WKB solution proportional to $\exp (i \int \sqrt{-V} dx)$ would be valid everywhere. This solution would be regular on the future horizon (for $\omega \le m \Omega_H$) and purely outgoing at infinity, i.e., the WKB approximation would give $Z_{\rm in}=0$. We would then have to examine whether this conclusion persisted beyond the WKB approximation. If so, we would have evidence of an instability. However, in Appendix \ref{app:WKB}, we show that this does not happen: the maximum of $P$ is positive in the range of interest $0 \le \omega \le m \Omega_H$.

Qualitatively, the form of $V$ is as follows. It takes the value $-(\omega-m\Omega_H)^2$ at the horizon, increases to a positive maximum and then decreases to $-\omega^2$ at infinity. In other words, there is a potential barrier of height proportional to $m^2$ separating the classically allowed region near infinity from the classically allowed region near the horizon. The WKB method will then give $|Z_{\rm out}/Z_{\rm in}| \approx 1$. However, as argued above, a mode at the threshold of instability will have $|Z_{\rm out}/Z_{\rm in}| \rightarrow \infty$. We conclude that no such mode exists for large $m$ and $\omega$.

The degenerate case in which $P$ vanishes at its maximum occurs only for extremal black holes with $\omega = m \Omega_H$. Equation (\ref{eqn:wronsk}) then gives $|Z_{\rm out}/Z_{\rm in}|=1$, so there is no evidence of any instability in this case either.

\subsection{Numerical results: asymptotically flat case}

We shall only consider gravitational perturbations in $D=7$ so $\mu=0$, $N=2$.

For given $l,m$, our strategy (following \cite{PT2}) is to start with small $\Omega_H$ and search the interval $0 \le \omega \le m \Omega_H$ for a solution of the radial equation that is regular on the future horizon and outgoing at infinity. This is then repeated for increasing values of $\Omega_H$ up to the maximum value.

To perform the numerical integration of the radial equation, we start by using the variable $\Phi$ defined by equation (\ref{eqn:Phidef}), which satisfies the equation
\be
 -  fg^{-1} \frac{d}{dr} \left( fg^{-1} \frac{d \Phi}{dr} \right) + 2i(\omega - m \Omega_H)fg^{-1} \frac{d\Phi}{dr} + \left[V + (\omega- m \Omega_H)^2 \right] \Phi = 0.
\ee
The advantage of working with $\Phi$ rather than $\Psi$ is that $\Phi$ is a smooth function of $r$ at $r=r_+$. Equation (\ref{eqn:Phibc}) determines the values of $\Phi$ and $\partial_r \Phi$ at $r=r_+$. The equation for $\Phi$ is integrated numerically out to $r=2r_+$, where we then calculate the corresponding values of $\Psi$ and $\Psi'$.\footnote{
In practice, we use the freedom to rescale $\Psi$ by a phase to set $\Psi=\Phi$ at $r=2r_+$.} We then take these as initial data for the integration of the equation for $\Psi$ out to large $r$.

A convenient object to consider is the ratio
\be
\label{eqn:ratio}
 Z \equiv \frac{|Z_{\rm out}|}{|Z_{\rm in}|} = \lim_{r \rightarrow \infty} \frac{|\partial_r \Psi + i \omega \Psi|}{|\partial_r \Psi - i \omega \Psi|}.
\ee
We are looking for $\omega$ for which this ratio diverges, corresponding to a purely outgoing solution.
We evaluate the right hand side at $r=p r_+$, $p=10,11,\ldots$. Denote these values by $Z_p$. The numerical integration is terminated when the sequence $Z_p$ has converged to an accuracy of $0.1 \%$. The solution is rejected if the Wronskian ${\rm Im} (\bar{\Psi} \partial_x \Psi)$ varies by more than $0.1 \%$ during the numerical integration.

In all cases we have examined, the qualitative form of the potential is the same as we found in the WKB analysis above, i.e., a potential barrier with a positive maximum separates the classically allowed regions near the horizon and far from the black hole. The corresponding behaviour of the solution $\Psi$ is: oscillation near the horizon, exponential growth in the potential barrier region and then oscillation out to infinity. If the potential barrier is large then this implies that the amplitude of oscillation far from the black hole will be large. However, we are looking for a mode with $Z_{\rm in}=0$. From (\ref{eqn:wronsk}), such a mode obeys
\be
 \lim_{r \rightarrow \infty} |\Psi|^2 = \frac{m \Omega_H - \omega}{\omega}.
\ee
Hence, unless $\omega$ is very small, such a mode will {\it not} have a large amplitude. Hence it seems very unlikely that we will find a suitable mode when the potential barrier is large. Phrasing the argument slightly differently, if $\Psi$ is large then $Z_{\rm in}$ and/or $Z_{\rm out}$ must be large compared with the left hand side of equation (\ref{eqn:wronsk}), which implies $Z \approx 1$. More physically, if the potential barrier is large then one expects almost perfect reflection and very little transmission, so the amplitude of $\Psi$ is much greater far from the black hole than near the horizon.

\begin{figure}[h!]
\centering{\psfig{file=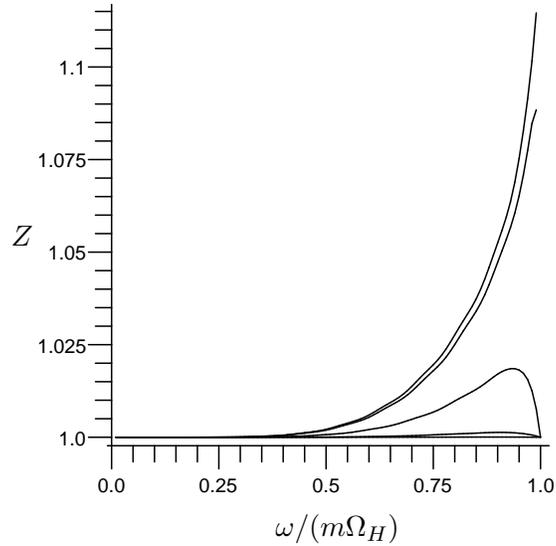,width=8cm}}
\begin{picture}(0,0)(230,0)
\put(10,120){$Z$}
\put(110,10){$\omega/(m\Omega_H)$}
\end{picture}
\caption{Plots of $Z$ againt $\omega/(m\Omega_H)$ for (from bottom to top) $\Omega_H/\Omega_{\rm max} = 0.5,0.7,0.9,0.99,0.999$ with $\epsilon=1,l=2,m=4$. Note that $Z=1$ for $\omega=m\Omega_H$ but this point has been deleted from the topmost two curves to make the figure clearer.}
\label{fig:ratio24}
\end{figure}

This argument suggests that we should examine the case for which the potential barrier is smallest. The potential barrier is minimized when $l$ is as small as possible and $m$ as large as possible. The most favourable case (using (\ref{eqn:mrange})) is therefore likely to be $l=2$, $\epsilon=1$, $m=4$. Our numerical results are shown in figure \ref{fig:ratio24}, where we plot $Z$ against $\omega /(m\Omega_H)$ for $\Omega_H/\Omega_{\rm max} = 0.5,0.7,0.9,0.99,0.999$ where $\Omega_{\rm max} = \sqrt{N/(N+1)}/r_+$ is the upper bound on $\Omega_H$. The curves have the same qualitative shape as for the Kerr black hole \cite{PT2}, i.e., $Z$ is very close to $1$ for small $\omega$, then increases to a maximum near $\omega=m\Omega_H$ and decreases back to $1$ at $\omega=m\Omega_H$ (the latter property follows from equation (\ref{eqn:wronsk})).
The position of the maximum tends towards $\omega=m\Omega_H$ as $\Omega_H \rightarrow \Omega_{\rm max}$.
The largest value for $Z$ is $Z = 1.115$ so there is no sign of $Z$ diverging anywhere, as would be required for an instability. Note that the amplification of energy flux in superradiant scattering is given by $Z^2$ so the maximum amplification apparent in our data is about $24 \%$, and is achieved as the black hole tends to extremality and $\omega \rightarrow m \Omega_H -$. This is just as for Kerr, although for Kerr, the maximum amplification is much greater: $138\%$  \cite{PT3}.

\begin{figure}[h!]
\centering{\psfig{file=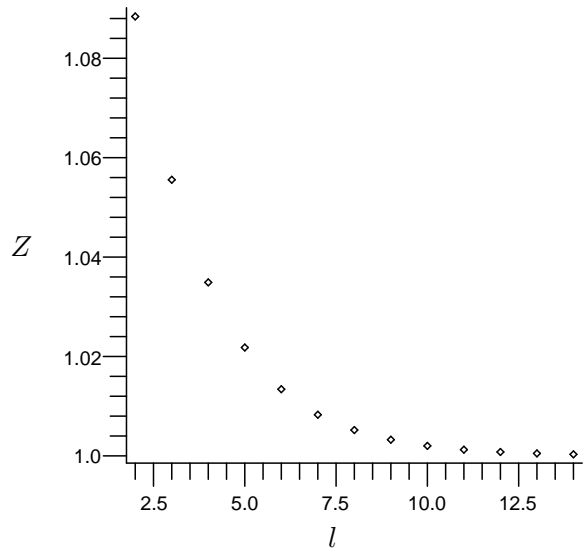,width=8cm}}
\begin{picture}(0,0)(240,0)
\put(10,120){$Z$}
\put(130,10){$l$}
\end{picture}
\caption{Plot of $Z$ against $l=2,3,\ldots$ for $m=m_{\rm max} = l+2$, $\epsilon=1$, $\Omega_H/\Omega_{\rm max}=0.99$, $\omega/(m\Omega_H) = 0.99$.}
\label{fig:ratiomdep}
\end{figure}

We have repeated our analysis for other values of $(l,m)$. The results
are qualitatively similar to the case we have just discussed. For
$\Omega_H/\Omega_{\rm max}=0.99$, the largest value of $Z$ obtained
for $\epsilon=1$ and $(l,m)=(3,5),(3,3)$ was $1.056,1.000$
respectively, reflecting the fact that decreasing $m$ tends to
increase the potential barrier. In figure \ref{fig:ratiomdep} we
exhibit how $Z$ varies with $l$ with $m=m_{\rm max}(l)=l+2$. The
largest value of $Z$ occurs for the $l=2$, $m=4$ case discussed above,
and $Z$ decreases monotonically to $1$ as $l$ increases, in agreement
with our WKB analysis.


\section{Discussion}

\label{sec:discuss}

We have shown that there exists a class of higher-dimensional rotating black hole solutions for which linearized gravitational perturbations can be studied analytically. We have concentrated on the particular case of doubly transverse tensor modes. The equations of motion for such modes decouple from other types of perturbation and, after separation of variables, reduce to a single ODE governing the behaviour in the radial direction. It will be interesting to see whether the equations of motion governing other types of gravitational perturbation can also be reduced to ODEs.

Our radial equation could be used to calculate quasi-normal modes of these black holes. This might be simpler than the calculation of quasi-normal modes of a Kerr black hole because for Kerr, the eigenvalues of the angular equation have to be determined numerically and then fed into the radial equation whereas for cohomogeneity-1 black holes, these eigenvalues (our $\lambda$) can be determined analytically.

In the asymptotically flat case, our WKB argument reveals that, for large $m$, the gravitational perturbations are stable. We have also investigated the $D=7$ ($N=2$) case numerically for smaller $m$, finding no sign of any instability. (A similar treatment for $D>7$, will need to answer two questions regarding charged doubly transverse tensor harmonics on $CP^N$ with $N>2$, namely what is the lower bound on $l$ and do hermitian ($\sigma=-1$) modes exist?) It would be interesting to repeat our analysis using the continued fraction technique \cite{leaver} as this may prove more efficient. It is also of obvious interest to investigate whether the analytic proof of mode stability for Kerr \cite{whiting} could be adapted to our radial equation.

Our proof that $m=0$ modes are stable is relevant to the conjectured existence \cite{HSR} of higher-dimensional rotating asymptotically flat black holes with fewer symmetries than any known solution. The motivation behind this conjecture is that a stationary black hole must be axisymmetric (which has now been proven for higher dimensional black holes \cite{HIW}), i.e., it must admit a rotational symmetry. However all known higher-dimensional black hole solutions have multiple rotational symmetries and therefore more symmetry than one would expect of a general stationary black hole. So perhaps there exist undiscovered solutions that, although axisymmetric, possess fewer symmetries than any known solution. If such solutions are continuously connected to the known solutions then one might expect to see evidence for them in perturbation theory as stationary ($\omega=0$) axisymmetric ($m=0$) zero modes separating stable low angular momentum black holes from unstable high angular momentum ones. However, we have seen that, for doubly transverse tensor perturbations, $m=0$ modes are stable and there are no stationary axisymmetric perturbations. Hence there is no evidence for the conjecture of \cite{HSR} in the sector of doubly transverse tensor perturbations. It will be interesting to see what happens for other types of perturbation.

We have studied the case of asymptotically anti-de Sitter black holes
in some detail. Our main result is that the superradiant instability
of such black holes discussed in \cite{HR} occurs precisely when the
angular velocity of the black hole exceeds the speed of light (in the
sense that the co-rotating Killing field becomes space-like on the
conformal boundary). In other words, the instability occurs precisely
when the stability argument of \cite{HR} fails. Our results for small
AdS black holes also enable us to address a loop-hole in the proof of
\cite{HR}, namely that it could be invalidated by the presence of a tachyonic scalar field (obeying the Breitenlohner-Freedman bound), which violates the dominant energy condition. For small black holes at least, such scalars do not behave any differently from more massive scalars, i.e., the threshold of instability is unaffected. It would be interesting to see whether the methods of \cite{HR} and \cite{BF} could be combined to give a more general proof of this.

Something that has not been addressed in the literature is the end-point of the instability of \cite{HR}. For the super-radiant instability of a Kerr black hole in the presence of a massive scalar, the evolution is clear: the black hole gradually loses energy and angular momentum to bound superradiant modes of the scalar field. These couple non-linearly to gravitational waves (and thereby to non-superradiant modes of the scalar field) so this energy and angular momentum is ultimately radiated to infinity. After a very long time the black hole will have lost all of its angular momentum this way.

In the AdS case, the evolution of the instability must be rather different.
A black hole that is initially rotating faster than light will lose energy and angular momentum into superradiant modes of all fields of the theory under consideration. However, in AdS this cannot be radiated to infinity so instead the energy and angular momentum in fields outside the black hole must accumulate over time and backreaction will become important. If the system ultimately settles down to an equilibrium state then this must be described by a new stationary, asymptotically AdS black hole solution. Presumably the angular velocity of this new solution will not exceed the speed of light. Note that this argument does not depend on the details of the theory: it would be valid if gravity were the only field present, so there should even exist new vacuum black hole solutions.

Further evidence for the existence of new solutions comes from our analysis of modes at the threshold of instability. Since these have $\omega = m \Omega_H$, they are preserved by the co-rotating Killing vector field $\xi$. However they break the symmetries generated by $\partial/\partial t$ and $\partial/\partial \psi$. The existence of these modes could indicate the presence of new branches of solutions that bifurcate from the solutions of \cite{HHT,GLPP} at the critical ($m$-dependent) value of $\Omega_H$. The new solutions would not be invariant with respect to $\partial/\partial t$ or $\partial/\partial \psi$. In other words, they would not be stationary or axisymmetric. However, $\xi$ would still describe a symmetry. Since the bifurcation point occurs when the original black hole is rotating faster than light, $\xi$ would be spacelike near infinity but timelike near the horizon. So, near the bifurcation point, the new branch of solutions would correspond to black holes that are not stationary in the usual sense but nevertheless admit a Killing field that is timelike near the horizon. This Killing field becomes spacelike {\it outside} an "ergosphere" (this is what happens for the co-rotating Killing field of the Kerr black hole). However, if such solutions do exist, they are themselves rotating faster than light and therefore seem very likely to be unstable.\footnote{
This is reminiscent of what happens for black strings. These suffer from the Gregory-Laflamme instability \cite{GL}. They admit a static zero mode \cite{GL2} that signifies the onset of instability \cite{HSR2} and suggests the existence of a new family of solutions bifurcating from the known ones \cite{HM}. This is indeed the case \cite{gubser,wiseman} but the new solutions have too low an entropy to be the endpoint of the original instability, and seem likely to be unstable themselves.}

What happens as one moves further away from a bifurcation point? Obviously we can only speculate, but one possibility is that, if one moves sufficiently far along one of the new branches of solutions, one reaches solutions for which $\xi$ is timelike everywhere outside the horizon. These would correspond to genuinely stationary black holes which are nevertheless nonaxisymmetric. There would be no violation of the theorem that a stationary black hole must be axisymmetric \cite{HIW} because this theorem assumes that the stationary Killing field is {\it not} normal to the event horizon whereas $\xi$ {\it is} normal to the horizon of all the black holes we have been discussing. If such black holes exist then it is natural to guess that these should be the new solutions describing the endpoint of the superradiant instability.

The possibility of a black hole being stationary with respect to a Killing field that does not approach the "usual" generator of global AdS time translations deserves further comment. Consider $AdS_5$ in global coordinates:
\be
 ds^2 = - \left( 1 + \frac{r^2}{\ell^2} \right) dt^2 + \left( 1 +   \frac{r^2}{\ell^2} \right)^{-1} dr^2 + r^2 \left( d\theta^2 + \sin^2 \theta d\phi_1^2 + \cos^2  \theta d\phi_2^2 \right).
\ee
The generator of global time translations $\partial/\partial t$ is a globally timelike Killing field whose norm diverges at infinity. Now consider the Killing field
\be
 \frac{\partial}{\partial t} + \frac{1}{\ell} \frac{\partial}{\partial \phi_1} +   \frac{1}{\ell} \frac{\partial}{\partial \phi_2}.
\ee
This is globally timelike with {\it constant norm}. The same construction works in any odd-dimensional AdS spacetime. Therefore there are (at least) two qualitatively different ways that an odd-dimensional asymptotically AdS space-time can be stationary: the generator of time-translations could have either unbounded norm or bounded norm. This does not appear to have been discussed before. The known AdS black hole solutions are stationary in both senses because they admit Killing fields that have the same asymptotic behaviour as $\partial/\partial t$ and $\partial/\partial \phi_i$ above. However there may well exist AdS black holes with less symmetry that are stationary only with respect to a Killing field of bounded norm.

We do not know whether the stationary nonaxisymmetric black holes discussed above must be of this form. If they are, then, since the stationary Killing field must be normal to the horizon, and since a Killing field of bounded norm is null on the conformal boundary, such black holes must be rotating at the speed of light.\footnote{
The terminology here may be a little confusing. If one defines angular velocity for such black holes in the usual way then it will vanish because the stationary Killing field is normal to the horizon. However, this is angular velocity defined with respect to a stationary bulk observer. Since the stationary Killing field is null on the conformal boundary, such an observer actually rotates at the speed of light with respect to the boundary and hence so does the black hole.}
Superficially, this makes sense because the superradiant instability "switches off" when the rotation of a black hole no longer exceeds the speed of light so one might expect the black hole to evolve to a final state rotating precisely at the speed of light. However, the evolution of the instability is a time-dependent process, during which the notion of angular velocity is not even defined, so we see no reason why the final time-independent state could not be rotating slower than light.

Finally, we note that supersymmetric black holes exist in $AdS_5$ \cite{susy}. Supersymmetry guarantees the existence of a non-spacelike Killing vector field that is normal to the event horizon \cite{HSR} and timelike with bounded norm at infinity \cite{GG}, i.e., these solutions have precisely the behaviour that we have just discussed and hence rotate at the speed of light. However, these solutions admit extra Killing fields analagous to $\partial/\partial \phi_i$ (and hence also $\partial/\partial t$) above so they are also stationary in the usual sense (i.e. with respect to a Killing field of unbounded norm). The existence of these extra Killing fields appears unrelated to supersymmetry, which raises the question of whether there exist more general supersymmetric black hole solutions without these extra symmetries. Such black holes would be nonaxisymmetric, and stationary only in the new sense that we have been discussing.

\bigskip

\centerline{\bf Acknowledgments}

\medskip

\noindent We thank Gary Horowitz for discussions and comments on a draft of this paper. JL would like to thank St. John's College, Oxford, for
financial support and DAMTP, Cambridge, for hospitality. HSR is a Royal Society University Research Fellow.

\bigskip

\appendix

\section{The Lichnerowicz equation}

\label{app:lich}

Introduce an orthonormal basis
\be
 e^0 = fdt, \qquad e^1 = g dr, \qquad e^2 = h \left( d\psi + A-\Omega dt \right), \qquad e^i = r \hat{e}^i_a dx^a,
\ee where $\hat{e}^i$ is a vielbein on the base. The dual basis
defined by $\langle e^\alpha, e_\beta \rangle = \delta^\alpha_\beta$
is: \be e_0 = f^{-1}\left(
\partial_t+\Omega
\partial_{\psi} \right), \qquad e_1 = g^{-1} \partial_r, \qquad e_2
= h^{-1}
\partial_{\psi}, \qquad e_i = r^{-1} (\hat{e}_i - \langle A,
\hat{e}_i \rangle
\partial_{\psi} )\ee
where $\hat{e}_i$ are the dual vectors to $\hat{e}^i$. The spin
connection is defined by ($\alpha,\beta$ are tangent space,
$\mu,\nu$ are curved indices) \be \omega_{\mu \alpha \beta} =
e^\nu_\alpha \nabla_\mu e_{\beta \nu} \ee and obeys \be
 d e^\alpha = -\omega^\alpha{}_\beta \wedge e^\beta.
\ee
We find
\bea
 \omega_{01} &=& - \frac{f'}{fg} e^0 - \frac{h \Omega'}{2fg} e^2, \qquad \omega_{02} = - \frac{h \Omega'}{2 fg} e^1, \qquad \omega_{0i} = 0, \nonumber \\
 \omega_{12} &=& \frac{h\Omega'}{2fg} e^0 - \frac{h'}{gh} e^2, \qquad \omega_{1i} = -\frac{1}{rg} e^i, \qquad \omega_{2i} = \frac{h}{r^2} J_{\hat{i} \hat{j}} e^j, \\
 \omega_{ij} &=& -\frac{h}{r^2} J_{\hat{i} \hat{j}} e^2 + \frac{1}{r} \hat{\omega}_{kij} e^k, \nonumber
\eea
where $\hat{i},\hat{j}$ denote components with respect to the base space vielbein, and $\hat{\omega}_{kij}$ are the components of the base space connection with respect to the base space vielbein.

The curvature two forms are given by $R^{\alpha}_{\phantom{d}\beta} = d\omega^{\alpha}_{\phantom{d}\beta} + \omega^{\alpha}_{\phantom{d}\gamma} \wedge \omega^{\gamma}_{\phantom{d}\beta}$. We find:
\begin{eqnarray}
R_{01} &=& \left[\frac{f'^2}{f^2g^2} + \frac{1}{g}\left(\frac{f'}{fg}\right)' - \frac{3h^2{\Omega'}^2}{4f^2g^2} \right] e^{0}\wedge e^{1} + \left[-\frac{1}{g}\left(\frac{h\Omega'}{2fg}\right)'- \frac{h'\Omega'}{fg^2} \right]e^{1}\wedge e^{2} - \frac{h^2\Omega'}{2fgr^2}J_{\hat{i}\hat{j}} e^{i}\wedge e^{j} \nonumber \\
R_{02} & = & \left[\frac{f'h'}{fhg^2} + \left(\frac{h\Omega'}{2fg} \right)^2 \right] e^{0} \wedge e^{2} \nonumber \\
R_{0i} & = & \frac{f'}{fg^2r} e^{0} \wedge e^{i} + \frac{h\Omega'}{2fg^2r} e^{2}\wedge e^{i} - \frac{h^2\Omega'}{2fgr^2} J_{\hat{i}\hat{j}} e^{1}\wedge e^{j} \nonumber \\
R_{12} & = & \left[-\frac{1}{g}\left(\frac{h\Omega'}{2fg}\right)' - \frac{h'\Omega'}{g^2f} \right] e^{0}\wedge e^{1} + \left[-\frac{1}{g}\left(\frac{h'}{gh} \right)' - \left( \frac{h'}{hg}\right)^2 - \left(\frac{h \Omega'}{2fg} \right)^2 \right ] e^{1}\wedge e^{2} \nonumber  \\ & + &\frac{1}{gr^2}\left(\frac{h}{r} - h'\right)J_{\hat{i}\hat{j}} e^{i} \wedge e^{j} \nonumber \\
R_{1i} & = & \frac{h^2\Omega'}{2fgr^2}J_{\hat{i}\hat{j}} e^{0} \wedge e^{j} + \left[-\frac{1}{g^2r^2} - \frac{1}{g}\left(\frac{1}{gr} \right)' \right] e^{1}\wedge e^{i} + \frac{1}{gr^2}\left(\frac{h}{r} - h' \right)J_{\hat{i}\hat{j}} e^{2} \wedge e^{j} \nonumber \\
R_{2i} & = & \frac{h\Omega'}{2rfg^2} e^{0}\wedge e^{i} + \left[\frac{1}{g} \left(\frac{h}{r^2}\right)' + \frac{h}{gr^3} \right] J_{\hat{i}\hat{j}}  e^{1} \wedge e^{j} + \left(-\frac{h'}{hrg^2} + \frac{h^2}{r^4} \right) e^{2} \wedge e^{i} \nonumber \\
R_{ij} & = & \hat{R}_{ij} - \frac{\Omega'h^2}{r^2gf}J_{\hat{i}\hat{j}} e^{0} \wedge e^{1} + \left[-\frac{h'}{r^2g} - \frac{1}{g} \left(\frac{h}{r^2}\right)' \right] J_{\hat{i}\hat{j}} e^{1}\wedge e^{2} \nonumber \\ & + & \left[-\frac{1}{r^2g^2} \delta^{i}_{k}\delta^{j}_{l} - \frac{h^2}{r^4} J_{\hat{i}\hat{j}}J_{\hat{k}\hat{l}} - \frac{h^2}{r^4} J_{\hat{i}\hat{k}}J_{\hat{j}\hat{l}} \right] e^{k} \wedge e^{l} \nonumber
\end{eqnarray}
The only non-zero components of $\Delta_L h$ are (using (\ref{transverse})):
\bea
\Delta_L h_{2i} &=& -\frac{2h}{r^5} J^{\hat{j} \hat{k}} \mathcal{D}_{\hat{k}}
h_{\hat{i}\hat{j}}, \\ \Delta_L h_{ij} &=& r^{-4}( -2 \hat{R}_{ikjl} h_{\hat{k}\hat{l}} -
\mathcal{D}^2 h_{\hat{i}\hat{j}}) -\frac{2im}{r^4}(
h_{\hat{i}\hat{k}}J_{\hat{k}\hat{j}} +J_{\hat{k}\hat{i}} h_{\hat{k}\hat{j}})
+\frac{4h^2}{r^6}J_{\hat{p}\hat{j}}J_{\hat{k}\hat{i}} h_{\hat{k}\hat{p}} \nonumber
\\ &+& \left[ -f^{-2}(\omega-m\Omega)^2 +\frac{m^2}{h^2}
   +\frac{4h^2}{r^4}- \left( \frac{D-3}{rg^2} +\frac{h'}{hg^2}+
  \frac{f'}{fg^2}\right) \frac{R'}{R} \right. \nonumber \\  &-& \left. \frac{1}{gR}\left(
  \frac{R'}{g}\right)' - 2\ell^{-2}(D-1) \right] \frac{h_{\hat{i}\hat{j}}}{r^2}.
\eea
Plugging this into the $2i$ component of the linearised Einstein equations leads to equation (\ref{Jtrans}) and the $ij$ component gives (\ref{einstein2}).

\section{Charged tensor harmonics on $CP^2$}

\label{app:CP2}

To calculate the spectrum of $\Delta_L^A$ acting on the space of
charged doubly transverse traceless tensors on $CP^2$ we will use
the method of~\cite{pope:82}. We will use the notation and conventions
of~\cite{pope:82} (with $e=1$) for ease of comparison and then
convert to our notation at the end. We consider a compact
Einstein K\"ahler manifold of complex dimension two, with K\"ahler
metric $g_{ab}$ which satisfies $R_{ab}=\Lambda g_{ab}$ and has K\"ahler
form $J$. A gauge field $A$ is defined by $dA= (\Lambda/2)J$.

In~\cite{pope:82} it is shown that a general symmetric
two-tensor can be decomposed as:
\be
Y_{ab} =
J_{a}^{\phantom{a}c}\omega^0_{cb}+
L_{a}^{\phantom{a}c}\omega^+_{cb}+K_{a}^{\phantom{a}c}\omega^-_{cb}
\ee
where $\omega^0$ and $\omega^{\pm}$ are uniquely determined anti
self-dual two forms (the K\"ahler form $J$ is taken to be
self-dual), and $L$ is a particular $(0,2)$ complex
self-dual two-form and $K=\bar{L}$. Explicitly, in terms of complex
vierbeins $z^m$ (so $g=z^m \otimes \bar{z}^m$), $L=-i \bar{z}^1
\wedge \bar{z}^2$ and $J= \frac{i}{2} (z^1 \wedge \bar{z}^1+ z^2
\wedge \bar{z}^2)$. The covariant exterior derivative on a $p$-form $\omega$
of charge $n$ (with respect to $A$) is $D=d-inA$ where $d$ is the ordinary exterior
derivative . The two-form $J$ is
uncharged, $L$ has charge $n=-2$, $K$ has charge $n=2$, and all three are
gauge covariantly constant (i.e. annihilated by
$\mathcal{D}=\nabla -inA$). One can split the operator into $(1,0)$ and $(0,1)$ parts as $D=D^++D^-$ where $D^{+} =
\partial -inA^+$,
$D^-= \bar{\partial} -inA^-$ and $A^+,A^-$ are the $(1,0)$ and $(0,1)$ parts of $A$.

Next we define the following Laplacians: \bea \Delta =-(\star D
\star D+D \star D \star), \qquad \Delta^{\pm} =-2(\star D^{\mp}
\star D^{\pm} + D^{\pm} \star D^{\mp} \star), \eea which reduce to
the Hodge de Rham and (anti)holomorphic Laplacians when acting on
uncharged forms. The fact that these three Laplacians are equal when
acting on uncharged forms is a well-known result for K\"ahler
manifolds. Interestingly, this fact is also true when acting on
charged two-forms.

As discussed in the main text, for doubly transverse tensors, one
may consider the hermitian and antihermitian cases separately.\footnote{
Note that this is the only point at which we refer to the doubly transverse condition. Once we have divided into hermitian and anti-hermitian perturbations, the doubly tranverse condition is a consequence of the transverse condition.} In
fact it turns out that hermitian doubly transverse tensor harmonics
do not exist on $CP^2$. This is easy to prove,
at least in the uncharged case. Any hermitian tensor can be written
as $Y_{ab}=J_{a}^{\phantom{a}c}\omega^0_{cb}$. Transversality
implies $D\omega^0=0$, which reduces to $d\omega^0=0$ in the
uncharged case. From the cohomology of $CP^2$ it follows that
$\omega^0$ must be proportional to $J$. But $J$ is self-dual and $\omega^0$ anti-self dual hence $\omega^0=0$.

For antihermitian modes $\omega^0=0$. Note that $L_{a}^{\phantom{a}c}\omega^+_{cb}$ and $K_{a}^{\phantom{a}c}\omega^-_{cb}$ are the $(0,2)$ and $(2,0)$ parts of $Y_{ab}$, and we can treat these separately. It is easy to show from the explicit
form of $L$ and $K$ that transversality of $Y_{ab}$ implies
$D^{\pm}\omega^{\pm}=0$. Let us focus on the $(0,2)$ part. Results for the
$(2,0)$ part follow from complex conjugation.
In~\cite{pope:82} the eigenvalue of such a two tensor was calculated
for $\omega^+$ with charge $n=2$, corresponding to $Y_{ab}$ of zero charge. We
will generalise this to $Y_{ab}$ of arbitrary charge $n$ and thus
$\omega^+$ of charge $n+2$ (since $L$ has charge $-2$). We find that (dropping the subscript on $\omega^+$ henceforth)
\be \Delta_L^A (L_{a}^{\phantom{a}c}\omega_{cb}) = L_{ac}
\left(\Delta + 2\Lambda\right) \omega_{cb} \ee which can be proven
by showing that \be \Delta \omega_{ab} = -\mathcal{D}^2\omega_{ab} +
\frac{4\Lambda}{3}\omega_{ab} \ee for an anti-self dual two form
$\omega$ of arbitrary charge. The uncharged case follows from the
Hodge de Rham Laplacian on two-forms. The charged case can be proven
explicitly by showing that $\Delta- \Delta_{HdR} =
-\mathcal{D}^2+\nabla^2$ when acting on an anti self-dual charged
two-form. For this computation one needs the explicit form of the
Riemann tensor for a K\"ahler metric of constant holomorphic
sectional curvature (which $CP^2$ is an example of!), which can be
found in~\cite{pope:82}.

Following~\cite{pope:82} we define a differential operator $E$ which
when acting on a scalar $\phi$ gives an anti-self dual two form in
the kernel of $D^+$, just as we require. Explicitly, \be E\phi = D^+
\star D^+ (L\phi). \ee This allows one to construct all such
eigentensors of the Lichnerowicz operator from scalar eigenfunctions
of the Laplacian, which is the technique used in~\cite{pope:82}. In
particular since we want $\omega$ to be of charge $n+2$ we can
always set $\omega = E\phi_{n+4}$ for some scalar field of charge
$n+4$. Next we need the eigenvalues of $E\phi_n$ (since this is a
two form we may choose any of the Laplacians). Using \bea \label{id}
\{ D^+, D^- \} = -\frac{in\Lambda}{2} J, \qquad \star(J \wedge \star
D^+L\phi) = -i \star D^+L\phi \eea one can show \be \label{prob2}
\Delta E\phi_n = \left[\lambda_n + 2(n-1) \Lambda \right] E\phi_n
\ee where $-\mathcal{D}^2\phi_n = \lambda_n \phi_n$. Note this
formula was proved in~\cite{pope:82} for $n=2,4$. Putting these
results together we find: \be \Delta_L^A
(L_{a}^{\phantom{a}c}\omega_{cb}) = \left[\lambda_{n+4}+
2\Lambda(n+4)\right] L_{a}^{\phantom{c}c}\omega_{cb} \ee where
$\omega =E\phi_{n+4}$. Thus to finish the computation of the
eigenvalues of these tensor harmonics we need to know the
eigenvalues of charged scalars. Fortunately this was done for $CP^2$
in~\cite{warner:82}, which uses the same conventions as we have. Thus
first we discuss how to go between the conventions of~\cite{pope:82}
and ours. Firstly, we work with $\Lambda=6$. Secondly our definition
of charge is with respect to a potential $A$ which satisfies
$J=\frac{1}{2}dA$. Thus the charge $n$ in~\cite{pope:82} is related
to our charge $m$ via $n=\frac{2m}{3}$. The results
of~\cite{warner:82} give the eigenvalues of charged scalars: \be
\lambda_m = 4\left[ k^2+(|m|+2)k+|m| \right] \ee where $k =0,1,2,\ldots$. We have
introduced the modulus sign to take care of charges of negative sign, which can be deduced from those of positive sign by complex
conjugation. Remembering to convert the charge we see that the
eigenvalues of antihermitian $(0,2)$ tensor harmonics of
charge $m$ are given by: \be \label{evalues} \lambda = \lambda_{m+6}
+8(m+6) = l(l+4)-m^2-4m+12 \ee where $l=2k+|m+6|$ and $k = 0,1,2,\ldots$.
It turns out that $l$ is the corresponding quantum number on $S^5$.
The eigenvalues for antihermitian $(2,0)$ tensors follow by
complex conjugation and thus take the same form with $m \rightarrow -m$.

Now we face a subtlety concerning the ranges of $k$. Obviously the
eigentensors must be non-vanishing so we need to examine when
$E\phi=0$. Fortunately~\cite{pope:82} has addressed this issue
and we can use the same technique. This involves introducing the
adjoint of $E$, which we will denote by $E^{\dagger}$, with respect
to the Hodge inner product and thus maps two forms to scalars. An
explicit expression for this adjoint is given in~\cite{pope:82}: \be
E^{\dagger} \omega = \star D^-(K \wedge \star D^- \omega). \ee Note that
$E\phi=0$ if and only if $E^{\dagger} E\phi
=0$. Using (\ref{id}) one can prove: \be E^{\dagger}E\phi_n =
-(\lambda_n+n\Lambda)\left[\lambda_n+2(n-1)\Lambda\right]\phi_n. \ee Note that
this was proven in~\cite{pope:82} only for $n=2,4$. Therefore
$E\phi_{n+4}=0$ if and only if: \be \lambda_{n+4} =-(n+4)\Lambda,
\qquad \textrm{or} \qquad \lambda_{n+4} = -2(n+3)\Lambda. \ee
Converting to our notation, we thus learn that one must exclude any
values of $k$ for which: \be \lambda_{m+6} =-4(m+6), \qquad
\textrm{or} \qquad \lambda_{m+6} = -4(2m+9). \ee This is easily done
and we find that one must exclude $k=0,1$ for $m \leq -6$ and $k=0$
for $m=-5$. Pleasingly this implies $l\geq 2$ as one would expect
for tensor harmonics. Once again, simply let $m \to -m$ to deduce what
happens in the holomorphic case.

It is important to check that the modes we have considered are not
pure gauge. We do this by computing the eigenvalues of pure gauge
tensor modes with respect to $\Delta_L$. A pure gauge perturbation
is of the form $h_{\mu\nu} = \nabla_{(\mu} \xi_{\nu )}$ on the
spacetime. Reducing to the base will give $Y_{ab} = D_{(a}
\xi_{b)}$. Using the gauge transverse and traceless conditions and
working in the conventions of~\cite{pope:82} gives: \bea \Delta_L^A
Y_{ab} = 2\Lambda Y_{ab} -2in\Lambda J_{(a}^{\phantom{(a}c}Y_{b)c}
+\frac{3in\Lambda}{4}( J_{a}^{\phantom{a}c}D_b\xi_c +
J_b^{\phantom{b}c}D_a \xi_c ).\eea Now we specialise to
antihermitian $Y_{ab}$, so $JY=i\epsilon Y$; this
can be achieved by setting $J\xi =i\epsilon \xi$. Then \be \Delta_L^A
Y_{ab} = \Lambda \left( 2+\epsilon \frac{n}{2} \right) Y_{ab}. \ee
Converting to our conventions implies that pure gauge eigenvalues
take the form: \be \lambda = 2\left( 6+\epsilon m \right). \ee By
comparing these to (\ref{evalues}), and taking the ranges of $k$
into account it is easy to show that the two sets of eigenvalues do
not overlap so we do not need to worry about pure gauge modes.

\section{Asymptotically flat case: WKB potential}

\label{app:WKB}

The maximum value of the function $P$ defined by (\ref{eqn:Pdef}) is
\be
 P_{\rm max} = \frac{1}{N+1} \left( \sqrt{\frac{N}{N+1}} \frac{m}{\omega} \right)^{2N} Q,
\ee
where
\be
 Q = 1 - \frac{N+1}{1-Y^2}\left[X^{N} \left(1 - \sqrt{\frac{N}{N+1}} XY \right) \right]^2,
\ee
with
\be
 Y \equiv r_+ \Omega_H \le \sqrt{\frac{N}{N+1}}, \qquad X \equiv \sqrt{\frac{N+1}{N}} \frac{r_+ \omega}{m} \le \sqrt{\frac{N+1}{N}} Y \le 1,
\ee
where we have used $\omega \le m \Omega_H$ and the upper bound on
$\Omega_H$. The expression in square brackets is positive and
monotonically increasing (as a function of $X$) for $X \le \sqrt{\frac{N+1}{N}} Y$ and hence
\be
 Q \ge 1 - (N+1) \left( \frac{N+1}{N} \right)^{N} (1-Y^2) Y^{2N}.
\ee
The expression on the RHS is monotonically decreasing for $Y \le \sqrt{\frac{N}{N+1}}$, from which we conclude $Q \ge 0$ and hence $P_{\rm max} \ge 0$. Equality occurs if, and only if, $\omega=m\Omega_H$ and $\Omega_H = \sqrt{\frac{N}{(N+1) r_+^2}}$ (i.e. the black hole is extremal).


\begin{thebibliography}{99}

\bibitem{MP}
R.~C.~Myers and M.~J.~Perry,
``Black Holes In Higher Dimensional Space-Times,''
Annals Phys.\  {\bf 172}, 304 (1986).

\bibitem{ER}
R.~Emparan and H.~S.~Reall,
``A rotating black ring in five dimensions,''
Phys.\ Rev.\ Lett.\  {\bf 88} (2002) 101101
[arXiv:hep-th/0110260].

\bibitem{EM}
R.~Emparan and R.~C.~Myers,
``Instability of ultra-spinning black holes,''
JHEP {\bf 0309}, 025 (2003)
[arXiv:hep-th/0308056].

\bibitem{HG}
G.~Gibbons and S.~A.~Hartnoll,
``A gravitational instability in higher dimensions,"
Phys.\ Rev.\ D {\bf 66} (2002) 064024
[arXiv:hep-th/0206202].

\bibitem{kodama:03a}
H.~Kodama and A.~Ishibashi,
``A master equation for gravitational perturbations of maximally  symmetric
black holes in higher dimensions,''
Prog.\ Theor.\ Phys.\  {\bf 110}, 701 (2003)
[arXiv:hep-th/0305147].

\bibitem{kodama:03}
A.~Ishibashi and H.~Kodama,
``Stability of higher-dimensional Schwarzschild black holes,''
Prog.\ Theor.\ Phys.\  {\bf 110}, 901 (2003)
[arXiv:hep-th/0305185].

\bibitem{adscft}
J.~M.~Maldacena,
``The large N limit of superconformal field theories and supergravity,''
Adv.\ Theor.\ Math.\ Phys.\  {\bf 2}, 231 (1998)
[Int.\ J.\ Theor.\ Phys.\  {\bf 38}, 1113 (1999)]
[arXiv:hep-th/9711200],
S.~S.~Gubser, I.~R.~Klebanov and A.~M.~Polyakov,
``Gauge theory correlators from non-critical string theory,''
Phys.\ Lett.\ B {\bf 428}, 105 (1998)
[arXiv:hep-th/9802109],
E.~Witten,
``Anti-de Sitter space and holography,''
Adv.\ Theor.\ Math.\ Phys.\  {\bf 2}, 253 (1998)
[arXiv:hep-th/9802150].

\bibitem{HHT}
S.~W.~Hawking, C.~J.~Hunter and M.~M.~Taylor-Robinson,
``Rotation and the AdS/CFT correspondence,''
Phys.\ Rev.\ D {\bf 59} (1999) 064005
[arXiv:hep-th/9811056].

\bibitem{GLPP}
  G.~W.~Gibbons, H.~Lu, D.~N.~Page and C.~N.~Pope,
  ``The general Kerr-de Sitter metrics in all dimensions,''
  J.\ Geom.\ Phys.\  {\bf 53} (2005) 49
  [arXiv:hep-th/0404008], ``Rotating black holes in higher dimensions with a cosmological constant,'' Phys.\ Rev.\ Lett.\  {\bf 93}, 171102 (2004)
  [arXiv:hep-th/0409155].

\bibitem{HR}
S.~W.~Hawking and H.~S.~Reall,
``Charged and rotating AdS black holes and their CFT duals,''
Phys.\ Rev.\ D {\bf 61}, 024014 (2000)
[arXiv:hep-th/9908109].

\bibitem{kerrinstab}
T.~Damour, N.~Deruelle and R.~Ruffini,
``On Quantum Resonances In Stationary Geometries,''
Lett.\ Nuovo Cim.\  {\bf 15} (1976) 257.

\bibitem{eardley}
T.~J.~M.~Zouros and D.~M.~Eardley,
``Instabilities Of Massive Scalar Perturbations Of A Rotating Black Hole,''
Annals Phys.\  {\bf 118} (1979) 139.

\bibitem{detweiler}
S.~Detweiler,
``Klein-Gordon Equation And Rotating Black Holes,''
Phys.\ Rev.\ D {\bf 22}, 2323 (1980).

\bibitem{cardosoa}
V.~Cardoso, O.~J.~C.~Dias, J.~P.~S.~Lemos and S.~Yoshida,
``The black hole bomb and superradiant instabilities,''
Phys.\ Rev.\ D {\bf 70}, 044039 (2004)
[Erratum-ibid.\ D {\bf 70}, 049903 (2004)]
[arXiv:hep-th/0404096].

\bibitem{cardoso}
V.~Cardoso and O.~J.~C.~Dias,
``Small Kerr-anti-de Sitter black holes are unstable,''
Phys.\ Rev.\ D {\bf 70}, 084011 (2004)
[arXiv:hep-th/0405006].

\bibitem{teukolsky}
S.~A.~Teukolsky,
``Rotating Black Holes - Separable Wave Equations For Gravitational And
Electromagnetic Perturbations,''
Phys.\ Rev.\ Lett.\  {\bf 29}, 1114 (1972),
``Perturbations Of A Rotating Black Hole. 1. Fundamental Equations For
Gravitational Electromagnetic, And Neutrino Field Perturbations,''
Astrophys.\ J.\  {\bf 185}, 635 (1973).

\bibitem{frolov}
V.~P.~Frolov and D.~Stojkovic,
``Quantum radiation from a 5-dimensional rotating black hole,''
Phys.\ Rev.\ D {\bf 67}, 084004 (2003)
[arXiv:gr-qc/0211055].

\bibitem{ida}
D.~Ida, Y.~Uchida and Y.~Morisawa,
``The scalar perturbation of the higher-dimensional rotating black holes,''
Phys.\ Rev.\ D {\bf 67}, 084019 (2003)
[arXiv:gr-qc/0212035].

\bibitem{page1}
  M.~Vasudevan, K.~A.~Stevens and D.~N.~Page,
  ``Particle motion and scalar field propagation in Myers-Perry black hole
  spacetimes in all dimensions,''
  Class.\ Quant.\ Grav.\  {\bf 22} (2005) 1469
  [arXiv:gr-qc/0407030].

\bibitem{page2}
  M.~Vasudevan, K.~A.~Stevens and D.~N.~Page,
  ``Separability of the Hamilton-Jacobi and Klein-Gordon equations in  Kerr-de
  Sitter metrics,''
  Class.\ Quant.\ Grav.\  {\bf 22} (2005) 339
  [arXiv:gr-qc/0405125],
  H.~K.~Kunduri and J.~Lucietti,
  ``Integrability and the Kerr-(A)dS black hole in five dimensions,''
  Phys.\ Rev.\ D {\bf 71} (2005) 104021
  [arXiv:hep-th/0502124],
  M.~Vasudevan and K.~A.~Stevens,
  ``Integrability of particle motion and scalar field propagation in
  Kerr-(anti) de Sitter black hole spacetimes in all dimensions,''
  Phys.\ Rev.\ D {\bf 72} (2005) 124008
  [arXiv:gr-qc/0507096].

\bibitem{KLR}
H.~K.~Kunduri, J.~Lucietti and H.~S.~Reall, in preparation.

\bibitem{GPP}
  G.~W.~Gibbons, M.~J.~Perry and C.~N.~Pope,
  ``The first law of thermodynamics for Kerr - anti-de Sitter black holes,''
  Class.\ Quant.\ Grav.\  {\bf 22} (2005) 1503
  [arXiv:hep-th/0408217].

\bibitem{wu:76}
T.~T.~Wu and C.~N.~Yang,
 ``Dirac Monopole Without Strings: Monopole Harmonics,''
Nucl.\ Phys.\ B {\bf 107} (1976) 365.

\bibitem{warner:82}
N.~P.~Warner,
``The Spectra Of Operators On Cp**N,''
Proc.\ Roy.\ Soc.\ Lond.\ A {\bf 383}, 217 (1982).

\bibitem{pope:00}
P.~Hoxha, R.~R.~Martinez-Acosta and C.~N.~Pope,
``Kaluza-Klein consistency, Killing vectors, and Kaehler spaces,''
Class.\ Quant.\ Grav.\  {\bf 17}, 4207 (2000)
[arXiv:hep-th/0005172].

\bibitem{pope:82}
C.~N.~Pope,
``Kahler Manifolds And Quantum Gravity,''
J.\ Phys.\ A {\bf 15}, 2455 (1982).

\bibitem{boucetta}
M. Boucetta, "Spectre du Laplacien de Lichnerowicz sur les projectifs complexes," C.R. Acad. Sci. Paris S\'er. I Math., {\bf 333}, 571 (2001).

\bibitem{morisawa}
Y.~Morisawa and D.~Ida,
``Scalar field perturbation on six-dimensional ultra-spinning black  holes,''
Phys.\ Rev.\ D {\bf 71}, 044022 (2005)
[arXiv:gr-qc/0412070].

\bibitem{cardoso2}
V.~Cardoso, G.~Siopsis and S.~Yoshida,
``Scalar perturbations of higher dimensional rotating and ultra-spinning
black holes,''
Phys.\ Rev.\ D {\bf 71}, 024019 (2005)
[arXiv:hep-th/0412138].

\bibitem{PT2}
W.H. Press and S.A. Teukolsky, "Perturbations of a rotating black hole. II. dynamical stability of the Kerr metric," Astrophys. J. {\bf 185}, 649 (1973).

\bibitem{hartle}
J.B. Hartle and D.C. Wilkins, "Analytic properties of the Teukolsky equation," Commun. Math. Phys. {\bf 38}, 47 (1974).

\bibitem{PT3}
S.A. Teukolsky and W.H. Press, "Perturbations of a rotating black hole. III. interaction of the hole with gravitational and electromagnetic radiation," Astrophys. J. {\bf 193}, 443 (1974).

\bibitem{BF}
P.~Breitenlohner and D.~Z.~Freedman,
``Stability In Gauged Extended Supergravity,''
Annals Phys.\  {\bf 144}, 249 (1982).

\bibitem{GM}
S.~S.~Gubser and I.~Mitra,
``The evolution of unstable black holes in anti-de Sitter space,''
JHEP {\bf 0108}, 018 (2001)
[arXiv:hep-th/0011127].

\bibitem{GL}
R.~Gregory and R.~Laflamme,
``Black strings and p-branes are unstable,''
Phys.\ Rev.\ Lett.\  {\bf 70}, 2837 (1993)
[arXiv:hep-th/9301052].

\bibitem{leaver}
E.~W.~Leaver,
``An Analytic Representation For The Quasi Normal Modes Of Kerr Black
Holes,''
Proc.\ Roy.\ Soc.\ Lond.\ A {\bf 402}, 285 (1985).

\bibitem{schutz}
B.F. Schutz and C.M. Will, "Black hole normal modes: a semianalytic approach," Astrophys. J. {\bf 291}, L33 (1985).

\bibitem{dicke}
R.H. Dicke and J.P. Wittke, "Introduction to quantum mechanics", Addison-Wesley, 1969.

\bibitem{cardoso3}
V.~Cardoso and S.~Yoshida,
``Superradiant instabilities of rotating black branes and strings,''
JHEP {\bf 0507}, 009 (2005)
[arXiv:hep-th/0502206].

\bibitem{whiting}
B.~F.~Whiting,
``Mode Stability Of The Kerr Black Hole,''
J.\ Math.\ Phys.\  {\bf 30}, 1301 (1989).

\bibitem{HSR}
H.~S.~Reall,
``Higher dimensional black holes and supersymmetry,''
Phys.\ Rev.\ D {\bf 68}, 024024 (2003)
[Erratum-ibid.\ D {\bf 70}, 089902 (2004)]
[arXiv:hep-th/0211290].

\bibitem{HIW}
S.~Hollands, A.~Ishibashi and R.~M.~Wald,
``A Higher Dimensional Stationary Rotating Black Hole Must be Axisymmetric,''
arXiv:gr-qc/0605106.

\bibitem{GL2}
R.~Gregory and R.~Laflamme,
``Hypercylindrical Black Holes,''
Phys.\ Rev.\ D {\bf 37}, 305 (1988).

\bibitem{HSR2}
H.~S.~Reall,
``Classical and thermodynamic stability of black branes,''
Phys.\ Rev.\ D {\bf 64}, 044005 (2001)
[arXiv:hep-th/0104071].

\bibitem{HM}
G.~T.~Horowitz and K.~Maeda,
``Fate of the black string instability,''
Phys.\ Rev.\ Lett.\  {\bf 87}, 131301 (2001)
[arXiv:hep-th/0105111].

\bibitem{gubser}
S.~S.~Gubser,
``On non-uniform black branes,''
Class.\ Quant.\ Grav.\  {\bf 19}, 4825 (2002)
[arXiv:hep-th/0110193].

\bibitem{wiseman}
T.~Wiseman,
``Static axisymmetric vacuum solutions and non-uniform black strings,''
Class.\ Quant.\ Grav.\  {\bf 20}, 1137 (2003)
[arXiv:hep-th/0209051].

\bibitem{susy}
J.~B.~Gutowski and H.~S.~Reall,
``Supersymmetric AdS(5) black holes,''
JHEP {\bf 0402}, 006 (2004)
[arXiv:hep-th/0401042],
``General supersymmetric AdS(5) black holes,''
JHEP {\bf 0404} (2004) 048
[arXiv:hep-th/0401129],
Z.~W.~Chong, M.~Cvetic, H.~Lu and C.~N.~Pope,
``Five-dimensional gauged supergravity black holes with independent  rotation
parameters,''
Phys.\ Rev.\ D {\bf 72}, 041901 (2005)
[arXiv:hep-th/0505112],
``General non-extremal rotating black holes in minimal five-dimensional
gauged supergravity,''
Phys.\ Rev.\ Lett.\  {\bf 95}, 161301 (2005)
[arXiv:hep-th/0506029],
H.~K.~Kunduri, J.~Lucietti and H.~S.~Reall,
``Supersymmetric multi-charge AdS(5) black holes,''
JHEP {\bf 0604}, 036 (2006)
[arXiv:hep-th/0601156].

\bibitem{GG}
J.~P.~Gauntlett and J.~B.~Gutowski,
``All supersymmetric solutions of minimal gauged supergravity in five
dimensions,''
Phys.\ Rev.\ D {\bf 68}, 105009 (2003)
[Erratum-ibid.\ D {\bf 70}, 089901 (2004)]
[arXiv:hep-th/0304064].



\end{thebibliography}
\end{document}